\documentclass[hidelinks,11pt]{article}
\usepackage{graphicx}
\usepackage{xcolor}
\usepackage{amssymb}
\usepackage{amsmath}
\usepackage{mathabx}
\usepackage{bbm}
\usepackage{multirow}
\usepackage{multicol}
\usepackage{amscd}
\usepackage{mfpic}
\usepackage{verbatim}
\usepackage{cancel}
\usepackage{chemarrow}
\usepackage[linktocpage=false,colorlinks = true,
            linkcolor = darkviolet,
            urlcolor  = blue,
            citecolor = blue,
            anchorcolor = magenta]{hyperref}

\definecolor{cream}{rgb}{.97, .95, .88}
\definecolor{darkcream}{rgb}{1., .88, .5}
\definecolor{lightpink}{rgb}{0.98, 0.88, 0.87}
\definecolor{lightwhite}{rgb}{1., 0.98, 0.95}
\definecolor{lightsalmon}{rgb}{1., 0.95, 0.90}
\definecolor{lightviolet}{rgb}{0.9, 0.8, 0.9}
\definecolor{lightgray}{rgb}{.96, .96, .96}  
\definecolor{lgray}{rgb}{.75, .75, .75}
\definecolor{LemonChiffon}{rgb}{0.95, 1., 0.7}
\definecolor{lightolivegreen}{rgb}{0.84, 0.89, 0.25}
\definecolor{lightgreen}{rgb}{.664, 1., .52}
\definecolor{llgreen}{rgb}{.900, .983, .960}
\definecolor{tristle}{rgb}{0.87, 0.67, 0.77} 
\definecolor{pink}{rgb}{0.95, 0.45, 0.75}
\definecolor{magenta}{rgb}{1., 0, 1.}
\definecolor{violet}{rgb}{0.9, 0.20, 0.85}
\definecolor{darkolivegreen}{rgb}{0.55, 0.65, 0.35}
\definecolor{maroon}{rgb}{0.7, 0.26, 0.56}
\definecolor{lightmaroon}{rgb}{0.85, 0.38, 0.58}
\definecolor{darkmaroon}{rgb}{0.604, 0.169, 0.451}
\definecolor{ddarkmaroon}{rgb}{0.2, 0.03125, 0.150}
\definecolor{mediumorchid}{rgb}{0.8, 0.33, 0.83}
\definecolor{mediumorchidd}{rgb}{1., 0.33, 0.63}
\definecolor{darkgreen}{rgb}{0.1, 0.6, 0.13}
\definecolor{lightyellow}{rgb}{1., 1., 0.82}
\definecolor{turquoise}{rgb}{0.042, 0.586, 0.512}
\definecolor{turquoisel}{rgb}{0.66, 0.94, 0.83}
\definecolor{darkturquoise}{rgb}{0.21, 0.55, 0.50}
\definecolor{coral}{rgb}{1., 0.6, 0.21}
\definecolor{lightorange}{rgb}{1., 0.88, 0.75}
\definecolor{orangered}{rgb}{1., 0.5, 0.}
\definecolor{orange}{rgb}{1., 0.65, 0.1}
\definecolor{orangel}{rgb}{1., .85, .3}
\definecolor{darkorange}{rgb}{0.875, 0.4, 0.204}
\definecolor{ddarkorange}{rgb}{.675, .218, .05}
\definecolor{bluesky}{rgb}{0.48, 0.53, 1.}
\definecolor{gold}{rgb}{1., 0.85, 0.25}
\definecolor{goldd}{rgb}{0.95, 0.75, 0.05}
\definecolor{darkviolet}{rgb}{0.54, 0.04, 0.84}
\definecolor{ddarkviolet}{rgb}{.382, .063, .657}
\definecolor{lightblue}{rgb}{0.30, 0.86, 0.89}
\definecolor{LightBlue}{rgb}{0.68, 0.85, 0.9}
\definecolor{lblue}{rgb}{0.78, 0.90, 0.95}
\definecolor{darkblue}{rgb}{.105, .308, .707}
\definecolor{lightmaroon}{rgb}{0.85, 0.38, 0.58}
\definecolor{darkmaroon}{rgb}{0.604, 0.169, 0.451}
\definecolor{darkpink}{rgb}{0.879, 0.020, 0.766}
\definecolor{ddarkpink}{rgb}{0.738, 0.195, 0.406}
\definecolor{grey}{rgb}{0.717, 0.717, 0.717}
\definecolor{lightgrey}{rgb}{0.800, 0.800, 0.800}
\definecolor{brown}{rgb}{0.740, 0.323, 0.182}
\definecolor{redbrown}{rgb}{.575, .158, .05}
\definecolor{darkbrown}{rgb}{0.34, 0.25, 0.05}
\definecolor{orangebrown}{rgb}{0.433, 0.262, 0.06}
\definecolor{pinkl}{rgb}{1., 0.788, 0.918}
\definecolor{salmon}{rgb}{1., 0.66, 0.5}
\definecolor{lightbrown}{rgb}{0.703, 0.508, 0.121}

\def\Journal#1#2#3#4{{#1} {\bf #2}, (#3) #4}

\def\etal{{\it et al.}}
\def\Name#1#2 {{#2, }{#1}}

\def\AIP{\em AIP Conf. Proc.}

\def\APH{\em Annals Phys.}

\def\ATM{\em Adv. Theor. Math. Phys}

\def\CQG{\em Class. Quant.~Grav.}

\def\FOP{\em Found. Phys.}

\def\GCS{\em Grav. Cosmol. Suppl.}
\def\GRG{\em Gen. Rel. Grav}

\def\IMA{{\em Int. J. Mod. Phys.} \emph{A}}
\def\IMD{{\em Int. J. Mod. Phys.} \emph{D}}

\def\JHE{\em J. High Energy Phys.}
\def\JMP{\em J. Math. Phys.}

\def\JPC{\em J. Phys. Conf. Ser.}

\def\JPU{\em J. Phys. (USSR)}

\def\JSM{\em J. Stat. Mech.}
\def\JSP{\em J. Stat. Phys.}
\def\LNP{\em Lect. Notes Phys.}

\def\MDU{\em MDPI Universe J.}

\def\MPL{{\em Mod. Phys. Lett.} \emph{A}}

\def\NAT{\em Nature}

\def\NCE{\em Nuovo Cimento}

\def\NPB{{\em Nucl.~Phys.}~\emph{B}}
\def\NPBS{{\em Nucl.~Phys.}~\emph{B}~Proc.Suppl}

\def\PLB{{\em Phys. Lett.}~\emph{B}}

\def\PRD{{\em Phys.~Rev.}~\emph{D}}
\def\PRL{\em Phys. Rev. Lett.}
\def\PRV{\em Phys.~Rev.}
\def\PRE{\em Phys.~Rep.}

\def\RMA{\em Rept.Math.Phys.} 
\def\RMP{\em Rep. Mod. Phys.}

\def\be{\begin{equation}}
\def\ee{\end{equation}}
\def\bea{\begin{eqnarray}}
\def\eea{\end{eqnarray}}
\def\bes{\begin{equation*}}
\def\ees{\end{equation*}}
\def\beas{\begin{eqnarray*}}
\def\eeas{\end{eqnarray*}}

\def\tr{\text{tr}}

\def\bm{\mathcal B}

\def\hm{\mathcal H}
\def\mm{\mathcal M}
\def\nm{\mathcal N}
\def\hA{\hat{A}}
\def\hH{\hat{H}}
\def\hJ{\hat{J}}
\def\hL{\hat{L}}

\def\hT{\hat{T}}

\def\sD{\cancel{D}}
\def\sqgr{$SU(\infty)$-QGR }
\def\suinf{{SU(\infty)}}
\def\suinfa{{su(\infty)}}

\begin{document}
\begin{center}
{\Large \bf Comparing Quantum Gravity Models: String Theory, Loop Quantum Gravity, and Entanglement gravity 
versus $\suinf$-QGR}\\
\end{center}

\begin{center}
Houri Ziaeepour$^{a,b}$\footnote{Email: {\tt houriziaeepour@gmail.com}} \\
\end{center}
$^a$Institut UTINAM, CNRS UMR 6213, Observatoire de Besan\c{c}on, Universit\'e de Franche Compt\'e, 
41 bis ave. de l'Observatoire, BP 1615, 25010 Besan\c{c}on, France \\
$^b$Mullard Space Science Laboratory, University College London, Holmbury St. Mary, 
GU5 6NT, Dorking, UK


\begin{abstract}
In a previous work~\cite{houriqmsymmgr} we proposed a new model for Quantum GRavity(QGR) 
and cosmology, dubbed \sqgr. One of the axioms of this model is that Hilbert spaces of the Universe 
and its subsystems represent $\suinf$ symmetry group. In this framework, the classical spacetime is 
interpreted as being the parameter space characterizing states of the $\suinf$ representing Hilbert 
spaces. Using quantum uncertainty relations, it is shown that the parameter space - the spacetime - 
has a 3+1 dimensional Lorentzian geometry. Here after a review of \sqgr, including the demonstration 
that its classical limit is Einstein gravity, we compare it with several QGR proposals, 
including: string and M-theories, loop quantum gravity and related models, and QGR proposals inspired 
by holographic principle and quantum entanglement. The purpose is to find their common and analogous 
features, even if they apparently seem to have different roles and interpretations. The hope is that 
such exercise gives a better understanding of gravity as a universal quantum force and clarifies the 
physical nature of the spacetime. We identify several common features among the studied models: 
importance of 2D structures; algebraic decomposition to tensor products; special role of $SU(2)$ group 
in their formulation; necessity of a quantum time as a relational observable. We discuss how these 
features can be considered as analogous in different models. We also show that they arise 
in \sqgr without fine-tuning, additional assumptions, or restrictions.
\end{abstract}
\pagebreak
\tableofcontents

\section{\bf Introduction and results}
Several fundamental questions about gravity and spacetime are not still answered by general relativity 
or by various attempts to find a consistent quantum description for gravitational interaction. The most 
daunting of these issues are the dimension of spacetime, which is usually considered to be the observed 
(3+1) without any explanation for its origin. Moreover, general relativity and Einstein gravity 
do not specify what is the nature of spacetime, except that it is curved in presence of matter and 
energy. Most QGR models treat spacetime as a physical entity, which despite being coupled to matter, has 
an independent existence. Indeed, often quantization of gravitational interaction, which is necessary in 
a Universe with quantum matter~\cite{grinconsist}, is interpreted as inevitability of a quantized 
spacetime. There are, however, multiple evidence against this conclusion:
\begin {itemize}
\item It is demonstrated~\cite{greos} that Einstein equation can be obtained from the second law of 
thermodynamics and holographic principle, that is the  proportionality of entropy inside a null 
(light-like) surface to its area rather than volume~\cite{hologprin,hologprin0,hologprin1,hologrev}. 
Holographic behaviour has been also observed in many-body systems with negligible 
gravity~\cite{condmatterentropy,condmatterentropy0}. These observations confirm the conclusion 
of~\cite{greos} that Einstein equation should be considered as equation of state. This interpretation and 
universality of gravitational interaction imply that what is perceived as {\it space} and its geometrical 
properties, such as distance and curvature, represent the state of its matter content. Thus, it seems that 
spacetime and matter are inseparable aspects of the same physical reality/entity.

\item Even without holographic principle the fact that energy-momentum tensor of matter - the 
source of gravitational interaction - depends on the spacetime metric means that spacetime and matter 
are more intertwined than, for instance, bosonic gauge fields and their matter source in Yang-Mills models.

\item In Quantum Field Theory (QFT) spacetime or its dual energy-momentum mode space (but not both at the 
same time) are used as {\it indices} to keep track of the continuum of matter and radiation. Giving the 
fact that in a quantum realm the classical vacuum - the apparently empty space between particles - can be 
described as a sea of virtual - off-shell - quantum states~\cite{hourivacuum}, means that we could 
completely neglect the {\it physical space} - the perceived 3-dimensional space. This were possible if we 
could identify, tag, and order all real and virtual particles, for instance by using the strength of their 
mutual quantum entanglement~\cite{qmwithouttime,qgrspaceless,qgrentangle,qgrentangle1,qgrentangle2} or 
interaction strength~\cite{qgrlocalqm}. In this view, the classical Einstein equation could be interpreted 
as an equation of state, which dynamically modifies parameter (index) space according to variation of 
interactions and entanglement between particles, and with respect to a relational quantum 
clock~\cite{qmtimepage}. 

\item It is useful to remind that in most QGR models the dimension of spacetime is considered as 
a parameter and little attempt is made to explain why it has the observed value.
\end {itemize}

In the last decade or so progress in quantum information theory has motivated construction of QGR models 
which are not based on the quantization of a classical theory. They are sometimes called 
{\it Quantum First} models in the literature~\cite{qgrlocalqm1}. In addition, progress in quantum 
information has highlighted the crucial role of the division of the Universe to parts - subsystems - and 
a proper mathematical definition for what can be considered as a {\it distinguishable} quantum (sub)system. 
This concept has special importance for gravity, because as much as we know from general relativity, it is 
a universal force, coupling everything to the rest of the Universe. Indeed, we will see later in this work 
that some QGR models struggle to find a naturally factorized - tensor product - Hilbert space in which each 
factor can be considered as presenting the Hilbert space of a subsystem. In QFTs without gravity subsystems 
are {\it particles} or their collections. A priori the same concept can be applied to QGR. However, in the 
strong coupling limit of QGR spacetime/gravity and {\it matter} may be indistinguishable. Therefore, it is 
necessary to have physically and mathematically well defined description of may be called a distinguishable 
subsystem of the Universe. We also remind that tensor product of subsystems is not only important for 
gravitational interaction, but also for the meaningful definition of locality, quantum clock, quantum 
information flow and relative entropy, renormalization flow, and holographic principle. None of these 
concepts would make sense without mathematical and physical notion of distinguishable subsystems.

In~\cite{houriqmsymmgr,houriqmsymmeos} we proposed a model for a quantum Universe which can be classified 
in {\it quantum first} category. Here we call it \sqgr. A brief review of this model is given in 
Sec. \ref{sec:sqgr}. It is a fundamentally quantum model, in the sense that its axioms come from quantum 
physics and its formulation is not a quantized version of a classical model. It does not include in its 
foundation, neither explicitly nor implicitly, a background spacetime or ingredients from Einstein general 
relativity, such as entropy-area relation. It is shown that both spacetime and Einstein equation emerge 
from quantum properties. The physical space is identified  with the space of indices parameterizing 
Hilbert and Fock spaces of the Universe and their subsystems. Einstein equation presents the projection 
of relational evolution of subsystems on the parameter space - the only observable when experiments do 
not have sufficient sensitivity to observe quantum field of gravitational interaction.

In other Quantum First proposals usually a background spacetime is implicitly present in their axioms. 
Examples of such models are those described in~\cite{qgrentangle,qgrlocalqm}. There is also 
implicit assumption of a physical space in models based on the holographic principle - a hypothesis 
inspired from semi-classical general relativity~\cite{bhentropy,hawkingrad} such 
as~\cite{qgrentangle1,qgrentangle2}. Indeed, it is obvious that holography without a geometrical space 
is meaningless. By contrast, in \sqgr the classical physical space and time genuinely emerge from 
quantum structure of the model and the assumption that any physical entity must be inside the Universe.

Quantum First QGRs and other modern approaches to QGR at first sight seem very different from each others. 
However, history of science is full of cases where seemingly different theories and interpretations were 
finally turned up to present the same physical concept viewed in different perspectives. The best example 
is Schr\"odinger's wave mechanics and Heisenberg's matrix mechanics approaches to quantum mechanics, which 
were later proved to be equivalent. For this reason, any new theory should look for what it has in common 
with other relevant models, and what new concepts or interpretations it is proposing. Such verification is 
particularly necessary for new QGR proposals, because QGR has been under intensive investigation for close 
to a century. Moreover, giving the fact that at present none of the proposals is fully satisfactory or 
has observational support, a better understanding of common aspects of different candidates may provide a 
direction and path to further developments, and eventually to the true model, unless all the proposals 
are completely irrelevant. 

In this work we compare \sqgr with some of popular approaches to QGR, namely: symplectic models, 
including Loop Quantum Gravity (LQG) and related models; string theory and its closely related matrix 
models (M-theory) and Anti-de Sitter-Conformal Field Theory (AdS/CFT) duality - more generally 
gauge-gravity duality; and models based on the holographic principle and quantum entanglement. Although 
our purpose is to find similarities and analogous features of among these models, this investigation also 
clarify their principle differences, which may be equally useful for further theoretical development, 
and eventually discriminating or constraining these models in experiments and observations.

We do not consider more traditional approaches such as canonical 
quantization~\cite{qgrcanonical,qgrcanonical0,qgrgeometrody} (see e.g.~\cite{qgrearlyhist,qgrgeometrodrev} 
for a review) and ADM 3+1 method~\cite{admgr}. After decades of research, it is now clear that they do not 
lead to a consistent and renormalizable theory. Other models omitted here are QGR models based on 
non-commutative spacetimes, models based on the quantum history interpretation of quantum mechanics, 
and the causal sets. These models are based on postulates that fundamentally deviate from those of 
\sqgr and their comparison with the latter is meaningless.

We begin by presenting a summary of the results of this work in the subsection \ref{sec:result}. Its 
purpose is to provide a quick overlook of comparison results. Therefore, if there are unclear points, 
the reader should refer to Sec. \ref{sec:comp} for more explanation. 

In Sec. \ref{sec:sqgr} and its subsections we briefly review \sqgr and show that the common features of 
QGR models arise naturally and without fine-tuning or addition of new assumptions to the initial 
axioms. Details of the comparison between models are discussed in Sec. \ref{sec:comp}. For each model 
we first briefly remind its main features. Then, we compare them with those of \sqgr. It 
is obvious that detailed and technical description of models and their variants, about which in some 
cases thousands of papers and numerous text books are written, is out of the scope of the present work. 
The aim of short reminds here is to introduce features and notations used for the comparison with \sqgr. 
Sec. \ref{sec:bgindep} reviews several background independent QGR models, including Ponzano-Regge model 
and LQG. Quantum First models are reviewed and compared with \sqgr in Sec. \ref{sec:qmqgr}. We compare 
string and M-theories, and gauge-gravity duality conjecture with \sqgr in Sec. \ref{sec:stringadscft}. 
A short outline is given in Sec. \ref{sec:outline}.

\subsection{\bf Summary of comparison results} \label{sec:result}
Form comparison of \sqgr proposal with some of other approaches to QGR we recognize a series of similar 
aspects, symmetries, and structures, which despite their different roles and interpretations in different 
models can be considered as analogous and common. Here we should emphasize that what we call 
{\it similarity} or {\it analogy} should not be interpreted as one-to-one correspondence. For instance, 
decomposition of $\suinf$ to $SU(2)$ factors in \sqgr is not the same operation as discretizing space 
to tetrahedra weighed by spins on their edges. Nonetheless, they have analogous mathematical 
descriptions - in this case a spin network. If the QGR models reviewed in this work contain at least some 
of the features and properties of the true theory, they should be, most probably, reflected in these 
common or analogous characteristics. 

The common features that we found in the models investigated in Sec. \ref{sec:comp} can be summarized 
as the followings:

\paragraph* {\bf Presence of 2-dimensional spaces or structures in the construction of models:}
\begin{description}
\item In some QGR models 2D spaces are used to construct a quantized space. They are either 2D boundary 
of a symplectic geometry consisting of connected tetrahedra with nonzero curvature at vertices, or  
2D worldsheets/membranes, embedded in a multi-dimensional space. These structures are treated as 
fundamental objects of the models - similar to {\it particles} in QFTs without gravity - and despite 
significant differences in their interpretation in different models, they have crucial role in 
generation of what is perceived as spacetime and gravity. In these models 2D structures are usually 
postulated and considered as physical entities. In this respect \sqgr is an exception, because 
diffeo-surfaces emerge from axioms and symmetries, and are considered as properties rather than being 
physical objects\footnote{Notice that the issue of what makes an abstract entity a {\it physical} object 
is rather philosophical. In practice, in mathematical formulations of physical phenomena all entities 
are abstract but related to what can be measured. Thus, in this sense they can be considered as 
{\it physical}.}. 

Extended nature of 2D structures has a crucial role in making QGR models renormalizable and in 
preventing singularities. In \sqgr this property is reflected in the fact that by definition a 
diffeo-surface cannot be shrunk to a point, otherwise $\suinf$ symmetry would be represented trivially. 
\end{description}

\paragraph* {\bf Decomposition to an algebraic tensor product:}
\begin{description}
\item Universe is a composite system and by definition, the Hilbert space of composite quantum systems 
is decomposed to tensor product of the Hilbert spaces of their subsystems~\cite{sysdiv}. Therefore, 
it is normal that an algebraic tensor product structure emerges, in one way or another, in the 
construction of QGR models. However, the most crucial tensor product structures in background 
independent models such as LQG and related models are spin-networks associated to the symplectic 
geometry and quantization of space. Specifically, their Hilbert space consists of all embedding of 
spin-weighted graphs - spin networks - generating the symplectic geometry states\footnote{The physical 
space can be considered as the state which is observed in the measurements.}. For this reason tensor 
products in spin networks cannot be interpreted as division to separable subsystems. This aspect is also 
shared by Quantum First models based on the entanglement.

In string and matrix theories, tensor products emerge in separation of compactified and non-compactified 
fields or as special configuration of string condensate in the form of D-branes (Moyal-Weyl solution). 
This can be interpreted as regarding spacetime and particles/matter fields as separate subsystems. 
Of course one may consider ensemble of strings, or more generally membranes or their presentation as 
large matrices as subsystems. However, they structures live in a higher dimensional space. In 
perturbative formulation this space has to be flat and it is not clear how strings/membranes interactions 
can generate it. In non-perturbative M-theory and matrix formulation, strings/membranes are {\it frozen} 
in a brane condensate in order to explain the observed (3+1)D spacetime. Their quantum fluctuations are 
treated similar to fields in QFTs, with {\it particles} as fundamental subsystems, and the $D=10$ 
dimensional background is static and unobservable.
\end{description}

\paragraph* {\bf ${SU(2)}$ group and spin network:}
\begin{description}
\item $SU(2)$ symmetry and/or its representations have a special role in most QGR models. In particular, 
they intervene in the construction of quantized geometry, because $SU(2) \cong SO(3)$ is the 
coordinate symmetry of the physical space.. Exceptions are string theory and \sqgr. Although 
$SU(2)$ group and its representations are extensively used in the formulation of \sqgr, it 
remains a purely mathematical utility without prior connection to the structure of classical spacetime.
\end{description}

\paragraph* {\bf A hidden or explicit $\suinf$ symmetry:}
\begin{description}
\item In models based on the symplectic construction of space the number of cells - usually tetrahedra - 
has to be considered to go to infinity to obtain a continuum at large distance scales - low energies. 
As these cells are indistinguishable from each others, the Hilbert space and dynamics of these models 
is invariant under $\suinf$ group defined on $\mathbb{R}$, rather than $\mathbb{C}$ considered in \sqgr.

String-gauge duality (M-theory) conjecture~\cite{stringgauge,stringgauge0,stringgauge1} identifies 
Yang-Mills models with large number of colors $N_c$ with string states. For $N_c \rightarrow \infty$ the 
symmetry of the Yang-Mills theory is $\suinf$. Indeed, in matrix model implementation of this conjecture, 
the fundamental objects are $N \times N, ~ N \rightarrow \infty$ matrices. and $SO(D)$ symmetry of 
fundamental $D=10$ dimensional space according to string theory can be interpreted as a special case of 
internal symmetries $G$ in \sqgr. However, despite these similarities, interpretation of matrices in 
M-theory and \sqgr are very different. Matrix models do not explore $\suinf$ symmetry and only use large 
matrices as representation of strings worldsheets or membranes in a special state of the fundamental 
background spacetime. By contrast, in \sqgr no special configuration is necessary to explain the observed 
(3+1)D spacetime. The tensor product $\suinf \times G$ provides mathematical requirements for division to 
subsystems~\cite{sysdiv} and grantees the existence of a perturbative expansion for both gravitational and 
matter sectors, without any constraint on the internal symmetry $G$. On the other hand, as 
$\suinf \times G \cong \suinf$, the model has also a non-perturbative limit.

\end{description}

\paragraph* {\bf Emergence of time and evolution as relative and relational phenomena:}
\begin{description}
\item A relational clock and its associated time parameter are necessary in the most QGR approaches 
except in string theory, in which time, space, and matter are treated in a same manner and are included
in the foundation of the model.
\end{description}

These common properties demonstrate that despite their apparent differences QGR candidates are more 
similar than probably expected. 

All the common features discussed here arise naturally and straightforwardly in \sqgr. Nonetheless, 
this model is a new proposal and much more must be done and understood about it before it can be 
considered as a genuine contender of a consistent and testable quantum gravity model. In particular, 
its predictions for the puzzle of black hole information loss and its predictions for future experiments 
seeking the detection of decoherence by quantum gravitational interactions should be investigated.

\section{\bf A brief review of \sqgr}  \label{sec:sqgr}
In this section we briefly summarize axioms, structure, and constituents of \sqgr. Only mathematical 
formulations necessary for the comparison with other QGR models are presented here.

\subsection{\bf Axioms and algebra} \label{sec:axiom} 
The \sqgr is based on 3 well motivated assumptions:
\setcounter{enumi}{0}
\renewcommand{\theenumi}{\arabic{enumi}}
\begin{enumerate} 
\item Quantum mechanics is valid at all scales and applies to every entity, including the Universe as 
a whole; \label{qmunivassum}
\item Any quantum system is described by its symmetries and its Hilbert space represents them; 
\label {symmassum}
\item The Universe\footnote{Here by Universe we mean everything causally or through its quantum 
correlation observable. Independent quantum observables correspond to mutually commuting hermitian 
operators applied to the Hilbert space. Their subspace is homomorphic to the Cartan subspace of the 
symmetry group of the quantum system, here the whole Universe ~\cite{houriqmsymm}.} has infinite 
number of independent degrees of freedom, that is mutually commuting observables. \label{infassum} 
\end{enumerate}

These axioms might seem trivial and generic. Here we briefly argue that they are not:
\begin{description}
\item Axiom \ref{qmunivassum} is not trivial because some QGR models extend or restrict quantum 
mechanics and/or QFT\footnote{We remind that QFT is not a model by its own. It is a specialized 
formulation of quantum mechanics suitable for description of many-body systems in spacetimes, specially 
in a Lorentz invariant manner.} in order to accommodate QGR, see Sec. \ref{sec:modqm} for a brief review 
of these models.
\item Axiom \ref{symmassum} is added to the above list because in postulates of quantum mechanics, as 
defined by Dirac~\cite{qmdirac} and von Neumann~\cite{qmvonneumann}, the Hilbert space is an abstract 
Banach space and no relation to symmetries is explicitly mentioned.  Axioms of quantum mechanics with 
symmetry as a foundational concept are described in~\cite{houriqmsymm}. Of course, in practice the Hilbert 
space is chosen such that it represent symmetries. However, this is due to the fact that the choice of 
Hilbert space for a quantum system is motivated by the configuration space of its classical limit and 
its symmetries. If we want to construct a fundamentally quantum model without referring to a 
corresponding classical system, we must specify how the Hilbert space is defined.
\item Axiom \ref{infassum} defines the symmetry of the system - the Universe, which as explained above 
is the basis for all other properties of the quantum system. Of course QFTs by definition have infinite 
number of observables/degrees of freedom - one or more at each point of the spacetime. However, in \sqgr 
there is no spacetime and the model is constructed as an abstract quantum system defined exclusively by 
its symmetry and its representation by the Hilbert space. 
\end{description}
On the hindsight, simplicity of these axioms is their advantage and in the following subsections we 
briefly review what can be concluded from this apparently generic assumptions. It should be reminded 
that we do not have any observed evidence of quantum gravity. Thus, sophisticated and {\it designed} 
axioms of some QGR models look rather imaginative, and one wonders why nature should have selected them 
among many other possibilities.

\subsection {\bf Representation of $\suinf$ group and Hilbert space} \label{sec:suinfrep}
The last assumption means that the Hilbert space of the Universe $\hm_U$ is infinite dimensional and 
represents $\suinf$ symmetry group, that is: 
\be
\bm[\hm_u] \cong \suinf  \label{buniv} 
\ee
where the sign $\cong$ means 
homomorphism and $\bm[\hm_u]$ is the space of bounded linear operators acting on $\hm_U$. 
Generators $\hL_{lm},~l \geqslant 0,~ |m| \leqslant l$ of $\bm[\hm_u]$ satisfy the Lie algebra: 
\be
[\hL_{lm},\hL_{l'm'}] = if_{lm,l'm'}^{l''m''} \hL_{l''m''} \label{statealgebra}
\ee
where structure coefficients $f_{lm,l'm'}^{l''m''}$ can be determined using properties of spherical 
harmonic functions, see e.g.~\cite{suninfhoppthesis} for more details. The reason for this property is 
that $\suinf$ can be decomposed to tensor products of $SU(2)$:
\be
\hL_{lm} = \mathcal {R} \sum_{i_\alpha = {1,~2,~3}, \alpha = {1, \cdots, l}} a^{(m)}_{i_1, \cdots i_l} \sigma_{i_1} \cdots 
\sigma_{i_l}, \quad \quad (l,m)~|~l = 0, \cdots, \infty; -l \leqslant m \leqslant +l \label {llmdef}
\ee
where $\sigma_{i_\alpha}$'s are $N \rightarrow \infty$ representations of Pauli 
matrices~\cite{suninfhoppthesis} and $\mathcal {R}$ is a normalization constant. Coefficients $a^{(m)}$ 
are determined from expansion of spherical harmonic functions with respect to spherical description of 
Cartesian coordinates~\cite{suninfhoppthesis}. 

The model is quantized using dual of its Hilbert space $\hm_U^*$ and its space of bounded linear 
operators $\bm[\hm_U^*]$:
\be
[\hL_a,\hJ_b] = -i \delta_{ab} \hbar, \quad \quad \hJ_a \in \bm[\hm_U^*] \label{lquantize}
\ee
$\hbar$ is the Planck constant.

It is known that $SU(\infty)$ is homomorphic to area preserving diffeomorphism of compact 2D 
surfaces~\cite{suninfhoppthesis,suninfym,suninftorus,suninfrep,suninfrep0}. From now on we use the 
shorthand name {\it diffeo-surface} for the surfaces which their area preserving diffeomorphism is 
homomorphic to $\suinf$ of interest. Diffeo-surfaces with different genus correspond to non-equivalent 
(non-isometric) representations of $\suinf$~\cite{suninfrep,suninfrep0}. These surfaces, and thereby 
$\bm[\hm_U] \cong \suinf$ are parameterized by two angular parameters $(\theta,~\phi)$. On the other hand, 
$\suinfa$ algebra is homomorphic to Poisson bracket of spherical harmonic functions, which for 
$\hbar = 1$ and dimensionless operators can be written as:
\bea
&& \hL_{lm} = i ~ \biggl (\frac{\partial Y_{lm}}{\partial \cos \theta} \frac{\partial}{\partial \phi} - 
\frac{\partial Y_{lm}}{\partial \phi} \frac{\partial}{\partial \cos \theta} \biggr ) = 
i ~ \sqrt{|g^{(2)}|} ~ \epsilon^{\mu\nu} (\partial_\mu Y_{lm}) \partial_\nu, 
\quad {\mu,~\nu \in \{\theta,\phi \}}\label{lharminicexp} \\
&& \hL_{lm} Y_{l'm'} = -i \{Y_{lm},~Y_{l'm'}\} = -i f ^{l"m"}_{lm,l'm'} Y_{l"m"} \label{lapp} \\
&& \{\mathsf{f},~\mathsf{g}\} \equiv \frac{\partial \mathsf{f}}{\partial \cos \theta} 
\frac{\partial \mathsf{g}}{\partial \phi} - \frac{\partial \mathsf{f}}{\partial \phi} 
\frac{\partial \mathsf{g}}{\partial \cos \theta}, \quad \forall ~ \mathsf{f},~\mathsf{g} \label{fgbrac}
\eea
In this representation of $\bm[\hm_U]$ vectors of the Hilbert space $\hm_U$ are complex functions of 
$(\theta,~\phi)$. If $\hL_{lm}$ (or equivalently $\hJ_{lm}$) operators are normalized by a constant factor 
proportional to $\frac{i\hbar}{cM_P}$, where $M_P$ is a mass scale - presumably Planck mass - the r.h.s. 
commutation relation (\ref{statealgebra}) becomes zero for $\hbar \rightarrow 0$ or 
$M_P \rightarrow \infty$ and the algebra of observables becomes Abelian, as in the classical mechanics. 
Thus, only when $\hbar \neq 0$ and $M_P < \infty$ the model presents a quantum system. This property 
establishes an inherent relationship between gravity and quantumness, as suggested in~\cite{houriqgr}. 

\subsubsection{\bf $\mathbf{SU(2)}$ in \sqgr}  \label{sec:su2}
The symmetry group $SU(2)$ has a special place in many QGR models, including in \sqgr where 
it is used for Cartan decomposition of $\suinf$ and description of its 
representations~\cite{suninfhoppthesis,suninfym,suninftorus,suninfrep,suninfrep0}. In particular, it 
allows to expand members of $\suinf$ as a linear function of spherical harmonic functions, 
analogous to an infinite spin chain. Consequently, generators of $\suinf$ are described by spin quantum 
numbers $(l,m)$. This representation is more suitable for practical applications than abstract complex 
functions of two angular parameters $(\theta,\phi)$. Nonetheless, one can easily transform one 
representation to the other, see e.g. appendices in~\cite{houriqmsymmgr}. 

We should emphasize that despite the importance of $SU(2)$ for \sqgr, it is not anything more than a 
mathematical tool. In fact, using the relation: 
\be
SU(N) \supseteq SU(N-K) \otimes SU(K)  \label{sundecompos}
\ee
the group $SU(\infty)$ can be decomposed to tensor products of any $SU(N),~ N < \infty$ by repeated 
application of (\ref{sundecompos}). Decomposition (\ref{llmdef}) corresponds to the case of $K=2$. It 
is the smallest non-Abelian special unitary group which can be used in the Cartan decomposition of 
$\suinf$.

\subsection{\bf Subsystems of the Universe}  \label{sec:subsys} 
In~\cite{houriqmsymmgr} it is shown that the quantum Universe as defined in the previous section is 
static and trivial. This is not a surprise, because there is no time parameter or a subsystem which 
plays the role of a quantum clock. On the other hand, according to a corollary in a description of 
quantum mechanics in which symmetry is considered to be foundational~\cite{houriqmsymm,houriqmsymm0}, 
this quantum system must inevitably be decomposable to subsystem. To this end, the Hilbert space must 
be factorized such that subsystems satisfy conditions defined in~\cite{sysdiv}. They include, among 
other things, factorization of the system's symmetry group and its representations. Using properties 
of $\suinf$~\cite{houriqmsymmeos}, in particular its multiplication~\cite{suninfrep0}:
\be
(\suinf)^n \cong \suinf \quad \forall ~ n > 0 \label{suinfn}
\ee
in~\cite{houriqmsymmgr,houriqmsymmeos} it is demonstrate that Hilbert spaces of subsystems have the 
general form of:
\be
\bm[\hm_s] \cong \suinf \times G  \label{sqgrsymm}
\ee 
where $\hm_s$ indicates the Hilbert space of a subsystem and $G$ is a finite rank symmetry group.
The presence of internal symmetries in the Standard model of particle physics is the main motivation 
for existence of $G$. Other motivations are discussed in~\cite{houriqmsymmgr}. Like any other quantum 
system, observables of a subsystem is defined by (\ref{sqgrsymm}) are hermitian members of 
$\bm[\hm_s]$. 

The Hilbert space of all subsystems is the tensor product of representations of the symmetry of 
subsystems (\ref{sqgrsymm}). Using (\ref{suinfn}), the Hilbert space of the ensemble of subsystems is:
\be
\bm[\bigotimes_s \hm_s] \cong \biggl (\suinf \times G \biggr)^{N \rightarrow \infty} \cong 
\suinf \times G^{N \rightarrow \infty}  \label{sqgrsymmall}
\ee
As $G^{N \rightarrow \infty} \cong \suinf$, (\ref{sqgrsymmall}) is consistent with (\ref{buniv}). Moreover, 
(\ref{sqgrsymmall}) shows that $\suinf$ factor of the Hilbert space can be considered as common to 
all subsystems. Thus, it has a role analogous to that of classical spacetime for all entities in the 
Universe. 

\subsubsection{\bf Parameter space of subsystems}  \label{secsubparam}
In addition to the emergence of an {\it internal} symmetry, the division of this quantum Universe 
induces a size or more precisely an area scale. Indeed, although the preserved area of one 
diffeo-surface is irrelevant for its diffeomorphism as representation of $\suinf$ group, it becomes 
important when parameter spaces of multiple systems with this symmetry, including the Universe as a 
whole, are compared. This is analogous to comparing finite intervals on a line with each others. 
An infinite line alone is scale invariant. But lengths of finite intervals can be compared with each 
others. This operation induces a length scale for the finite intervals and thereby for the whole line. 
Therefore, after division to subsystems the parameter space of $\suinf$ part of the Hilbert spaces of 
subsystems will depend on a third dimensionful parameter that we call $r$. It is measured with respect 
to a reference subsystem. Diffeo-surfaces of subsystems can be considered to be embedded in this 3D 
space. Notice that quantum state of a subsystem does not necessarily have a fixed $r$, and can be a 
superposition of pointer states with fixed $r$. 

Finally, to make the above setup dynamical, a relational dynamics and evolution 
\`a la Page \& Wootter~\cite{qmtimepage} or similar methods, see e.g.~\cite{qmtimedef} for a review, 
can be introduced by selecting one of the subsystems as a quantum clock. Variation of states of other 
subsystems are compared with the variation of state of the clock and is parameterized by a time parameter 
$t$. We interpret this 4D parameter space, which is homomorphic to $\mathbb{R}^{(4)}$ as the classical 
spacetime shared by all subsystems of the Universe. Of course the Hilbert space of every subsystem also 
has a factor representing its internal symmetry $G$, as shown in eq. (\ref{sqgrsymm}). As $\suinf$ and 
$G$ are considered to be orthogonal, their parameter spaces and actions on the states are separable. 
Specifically, $G$ transformations are performed {\it locally} to states $|t, r, \theta, \phi\rangle$, 
similar to a Yang-Mills gauge field defined on the classical spacetime. Due to this analogy, we identify 
the parameter space of the $\suinf$ symmetry with the classical spacetime.

\subsection{\bf Relation to classical geometry and Einstein equation}  \label{sec:geom}
Using Mandalestam-Tam uncertainty relation~\cite{qmspeed}, a quantity proportional to quantum fidelity 
of two close states $\rho$ and $\rho_1 = \rho + d\rho$ of subsystems (except reference and clock) can be 
defined~\cite{houriqmsymmgr}:
\be
ds^2 \equiv Q(\hH, \rho) dt^2 = \tr (\sqrt{d\rho} \sqrt{d\rho}^\dagger), \quad \quad 
Q(\rho, \hH) \equiv \frac{1}{2} |\tr([\sqrt{\rho}, \hH]^2)| \label{separation}
\ee
where $\hH$ is a Hamiltonian operator that generates the evolution of subsystems for the selected 
quantum clock associated to the time parameter $t$. Notice that here we have assumed that internal 
symmetry states of $\rho$ and $\rho_1$ are the same. We also remind that integrating out reference and 
clock subsystems makes state of other subsystem mixed and they should be treated as open quantum 
systems~\cite{qmrefsubsys}. 

The infinitesimal quantity $ds$ is a scalar of both the Hilbert space of subsystems and its parameter 
space. Due to the similarity of $ds$ to affine separation in Riemann geometry in the rest frame 
of subsystems, we can identify the two quantities up to an irrelevant normalization constant. Then, in an 
arbitrary reference frame of the parameter space $ds$ can be expanded as: 
\be
ds^2 = g_{\mu\nu} dx^\mu dx^\nu  \label{metric}
\ee
where $x^\mu$ is a point in the parameter space and $g_{\mu\nu}$ is the local metric. Using Mandalestam-Tam 
inequality, in~\cite{houriqmsymmgr} it is proved that the signature of metric $g_{\mu\nu}$ of the parameter 
space must be negative. Notice that the presence of a trace operator in the r.h.s. of (\ref{separation}) 
means that its l.h.s. is independent of the reference frame of the parameter space. This can be proved 
by expanding operators $\rho$ and $d\rho$ in an arbitrary basis $|t, r, \theta, \phi\rangle$ of the 
Hilbert space and calculating the trace in (\ref{separation}). Tracing amounts to integration over 
parameters $(t, r, \theta, \phi)$. Thus, $ds$ is independent of spacetime parametrization and coordinates 
$x^\mu$ in (\ref{metric}) should be considered as representative or average parameters of the quantum 
state $\rho$. In general relativity integration over the affine displacement $ds$ generates the world 
line of the system in the spacetime. Quantum systems do not follow a path in the classical phase space. 
Nonetheless, the world line generated by integration of $ds$ defined in (\ref{separation}) or 
(\ref{metric}) for the quantum subsystem can be interpreted as projection of the averaged path of the 
state in the Hilbert space into its parameter space - the spacetime.

These properties of \sqgr show that we have to find a pure quantum definition for extreme objects such 
as black holes, because their general relativity definition through their metric is highly degenerate 
and does not define their quantum state. Moreover, spacetime singularities of classical black holes 
may be irrelevant when they are considered as a many-body quantum state. Equations 
(\ref{separation}-\ref{metric}) are obtained from uncertainty principle~\cite{houriqmsymmgr}. Thus, 
similar to quantum mechanics, singularity of metric (\ref{metric}) can be interpreted as 
indistinguishability of states, or in other words infinite uncertainty. In any case, this important 
topic needs more investigation, once a proper quantum definition of a black hole is found.

Notice that here $t$ and $r$ are considered as classical values. More generally measurements preformed 
on the clock and on the subsystems relative to a reference to determine $r$ do not need to be projective. 
Such cases are intensively studied in the literature for general quantum 
systems~\cite{qmtimedef,qmrefsubsys} and we leave their application to \sqgr to future works.

\subsubsection{\bf Lorentz invariance of the parameter space}  \label{sec:lorentz}
A remark about Lorentz invariance of the parameter space is in order. In (\ref{metric}) this property 
is manifest. But, given the different origins of time, distance, and angular coordinates in \sqgr and 
their interpretation as classical spacetime, the question arises whether their $\mathbb{R}^{(3+1)}$ 
space is Lorentz invariant. The answer to this question is positive for following reasons:

\begin{description}
\item{- } Choices of a quantum clock and a reference subsystem for comparison between diffeo-surfaces 
are arbitrary. Change of these choices amount to changing corresponding parameters.
\item{- } Division to subsystems is not rigid and may change with change of clock and reference 
subsystem such that they respect necessary conditions defined in~\cite{sysdiv}. Thus, changing 
$t$ and $r$ in general lead to modification of $\suinf$ parameters $(\theta, \phi)$ and each of the new 
parameters $(t', r', \theta', \phi')$ would be a function of old parameters $(t, r, \theta, \phi)$.
\item{- } By definition the ensemble of subsystems must generate the static 2D Universe irrespective of 
how subsystems are defined and parameterized. This condition imposes Lorentz and diffeomorphism 
invariance on the parameter space - the spacetime.
\end{description}

\subsection{\bf Evolution} \label{sec:qmevol}
In this section we briefly review dynamics of the Universe as a whole and after its division to 
subsystems. We begin by constructing a symmetry-invariant functional for the whole Universe and then 
its modification when subsystems are taken into account.

\subsubsection{\bf The whole Universe} \label{sec:evolwhole}
According to symmetry description of quantum mechanics foundation~\cite{houriqmsymm}, the Universe as a 
whole is static and in a sort of equilibrium state. Specifically, using properties of $SU(N)$ groups, 
a $\suinf$-invariant functional consisting of elements of $\hm_U$ and $\bm[\hm_U]$ has the following 
form\footnote{More generally trace of multiplications of any number of generators of $SU(N)$ is 
invariant under $SU(N)$ transformation. However, their values are not independent and can be determined 
from structure coefficients. Using only the lowest order nonzero trace makes (\ref{twodlagrang}) 
equivalent to classical Lagrangian in QFT, despite the fact that the model does not come from a 
classical one.}:
\bea
\mathcal {L}_U & = & \int d^2\Omega  \biggl [\frac{1}{2} \sum_{a,~b} L^*_a(\theta,\phi) 
L_b (\theta, \phi) \tr (\hL_a \hL_b ) + \frac {1}{2} \sum_a \biggl (L_a (\theta, \phi) \tr (\hL_a \rho) + 
C.C \biggr ) \biggr ], \nonumber \\
&& d^2\Omega \equiv \sqrt{|g^{(2)}|} d(\cos \theta) d\phi \label{twodlagrang}
\eea
where $a = (l,m)$ or $(\theta', \phi')$, as explained in~\cite{houriqmsymmgr}. C-number amplitude  
$L_a$ determines the contribution of $\suinf$ generator $\hL^a$. We remind that the integration over 
angular coordinates of diffeo-surface is part of tracing operation, because generators 
(\ref{lharminicexp}) of the $\suinf$ symmetry are defined at each point of the diffeo-surface. By 
definition the whole Universe is in a pure state, because there is noting nothing outside, which could 
have been possibly traced out. Therefore, its density matrix can be written as 
$\rho = |\Psi \rangle\langle \Psi|$. In~\cite{houriqmsymmgr} it is explicitly shown that, as expected, 
applying variational principle with respect to amplitudes $L_a$ and components of the density matrix 
$\rho$ leads to a vacuum state as the equilibrium solution.

The action $\mathcal {L}_U$ is a formal description. In particular, it does not clarify how amplitudes  
$L_a$'s and density operator $\rho$ change under application of $\suinf$ group and reparametrization of 
diffeo-surface to preserve $\mathcal {L}_U$. This subject is described in details 
in ~\cite{houriqmsymmeos}, and as it is very important for the interpretation of the model as QGR and 
establishment of its relation with classical gravity, here we review the findings in some extent. 

We first remind that the surface element $d^2\Omega$ in (\ref{twodlagrang}) is invariant under 
reparametrization of angular coordinates. Thus, each term in the integrand must be reparametrization 
invariant. Moreover, for $SU(N)$ groups $\tr (\hL_a \hL_b) = C_a \delta_{ab}$, where $C_a$ is a constant. 
Therefore, phases of amplitudes $L_a$ in the first term of (\ref{twodlagrang}) are irrelevant. 
In addition, in the second term $\rho$ is hermitian and without loss of generality generators $\hL_a$ 
can be chosen to be hermitian too. Thus, phases of $L_a$'s are irrelevant and $L_a$'s can be considered 
to be real {\it fields} defined on the 2D diffeo-surface. 

Amplitudes $L_a$ must be invariant under translation 
$\theta \rightarrow \theta + \theta_0,~\phi \rightarrow \phi + \phi_0$ for arbitrary constant shift of 
the coordinates origin by $\theta_0$ and $\phi_0$, and rigid rotation of the frame. This means that $L_a$ 
must have a differential form with respect to coordinates $(\theta, \phi)$. Considering, in addition, the 
invariance under non-commutative $\suinf$ symmetry, we find that the first term in (\ref{twodlagrang}) 
should have the form of a 2D Yang-Mills Lagrangian for $\suinf$. Thus, $\mathcal {L}_U$ can written 
as~\cite{houriqmsymmgr}:
\bea
\mathcal {L}_U & = & \int d^2\Omega ~ \biggl [ ~\frac{1}{2} ~ \tr (F^{\mu\nu} F_{\mu\nu}) + 
\frac {1}{2} \tr (\sD \rho) \biggr ], \quad \quad \mu, \nu \in {\cos \theta, \phi} \label{yminvar} \\
F_{\mu\nu} & \equiv & F_{\mu\nu}^a \hL^a \equiv [D_\mu, D_\nu], \quad 
D_\mu = (\partial_\mu - \Gamma_\mu) \mathbbm{1} - \sum_a A_\mu^a \hL^a, \label{yminvardef} \\
F_{\mu\nu}^a F^{\mu\nu}_a & = & L^*_a L^a, \quad \forall a. \label{ltof} 
\eea
where $D_\mu$ is 2D covariant and gauge preserving derivative with an appropriate 2D connection 
$\Gamma_\mu$ (in $F^{\mu\nu}$ the connection will be canceled). Exact expression of the differential 
operator $\sD$ depends on the representation of 2D Euclidean group by the state $|\Psi\rangle$. For a 
scalar-type state $\sD = \overleftarrow {D}_\mu \overrightarrow {D}^\mu$ and for a spinor-type state 
$\sD = i \sigma^0 \sigma^i e_i^\mu \overleftrightarrow {D}_\mu$, where $\sigma^i ~ i= \{1,2\}$ are two of 
the $N \rightarrow \infty$-dimensional representation of Pauli matrices and $\sigma^0$ is the third 
Pauli matrix; $e_i^\mu$'s are zweibeins (analogous to vierbein in 2D). We remind that in 2D spaces 
the 2-form $F^a_{\mu\nu}$ has only one independent nonzero component. Therefore, the number of degrees 
of freedom in the two sides of (\ref{ltof}) is the same.

In Yang-Mills models the field strength $F_{\mu\nu}$ is a gauge invariant measurable. Moreover, in 
(\ref{yminvar}) variation of $L^a$'s and thereby $F_{\mu\nu}$ can be compensated by a diffeomorphism 
transformation of the compact 2D surface i.e the variation of $g^{\mu\nu}$. On the other hand, up to a 
global rescaling of the area of the diffeo-surface this transformation can be considered as application 
of $\suinf$ under which $F_{\mu\nu}$ is invariant. As we discussed before, the area of diffeo-surface 
of the whole Universe is not measurable. In this sense the first term in (\ref{yminvar}) is topological 
and can be identified, up to an irrelevant normalization constant, with the Euler class of the compact 
2D diffeo-surface: 
\be
\int d^2\Omega ~ \tr (F^{\mu\nu} F_{\mu\nu}) \propto \int d^2\Omega ~ \mathcal{R}^{(2)}  \label{eulerint}
\ee
where $\mathcal{R}^{(2)}$ is the 2D Ricci scalar of the parameter space - the diffeo-surface. This result 
is not surprising because a single indivisible quantum system is trivial~\cite{houriqmsymm}. In a 
geometrical view, if {\it local} details of the Universe are not distinguishable, only its global 
- topological - properties may characterize its states. In the present model the relevant global 
property is the topology of the diffeo-surface, corresponding to non-homomorphic representations of 
the $\suinf$ group~\cite{suninfrep,suninfrep0}. 

Equation (\ref{eulerint}) establishes the relation between \sqgr and classical gravity. Specifically, 
it shows that if quantum operators $F_{\mu\nu}^a \hL^a$ cannot be distinguished or observed, their 
overall effects is observed as variation of geometry of the parameter space. We can also interpret the 
r.h.s. of (\ref{eulerint}) as the projection of {\it dynamics} of the quantum Universe onto its 
parameter space.

\subsubsection{\bf Evolution of subsystems} \label{secevolsub}
When the Universe is divided to subsystems and a reference subsystem and a quantum clock are selected, 
it is still possible to construct a $\suinf$ invariant action functional similar to (\ref{yminvar}). 
It will depend on two additional parameters $r$ and $t$. They reflect the fact that $\suinf$ symmetry 
is now respected not only by the whole Universe, but also by its subsystems, which have acquired a new 
relative observable $r$ with respected to a selected reference subsystem, and their relative evolution 
is measured by a clock parameter $t$ with respect to a selected quantum clock. In addition, a full 
action must include terms invariant under the internal symmetry group of subsystems $G$. The formal 
description of this functional is~\cite{houriqmsymmgr}\footnote{Normalization of (\ref{lagrang}) is 
different from the expression in~\cite{houriqmsymmgr}. In the present normalization amplitudes $L_{lm}$ 
and $T_a$ are dimensionless}:
\bea
\mathcal {L}_{U_s} & = & \frac{1}{4\pi L_P^4} \int d^4 x \sqrt {-g} ~ \biggl [~ \frac {1}{4} ~ \biggl ( 
\sum_{l,m,l',m'} \tr (L^*_{lm} (x) L_{l'm'}(x) \hL_{lm} \hL_{l'm'}) + \nonumber \\
&& \sum_{l,m,a} \tr (L_{lm}(x) T_a (x) \hL_{lm} \otimes \hT_a) + 
\sum_{lm} L_{lm} ~ \tr (\hL_{lm} \otimes {\mathbbm 1}_G ~ \rho_s (x)) + \nonumber \\
&& \sum_{a,b} \tr (T^*_a (x) T_b (x) \hT_a \hT_b)\biggr ) + 
\frac{1}{2} \sum_a T_a ~ \tr ({\mathbbm 1}_{SU(\infty)} \otimes \hT_a ~ \rho_s (x)) \biggr ]. \label{lagrang}
\eea
where $T^a$'s are generators of the finite rank internal symmetry of subsystems $G$, and  
$L_P \equiv \sqrt{\hbar G_N / c^2 }$ is the Planck length. Notice that we have used Cartesian frame for 
coordinates of the 4D parameter space - the spacetime - and explicitly shown dimensionful coupling 
constant of $\suinf$ symmetry. Following the same line of arguments given for (\ref{twodlagrang}) about 
the invariance of parameter space under coordinate transformations, invariance under $\suinf$ 
transformations, and demonstration that the resulting action has the form of a Yang-Mills model with 
$\suinf$ symmetry, we conclude that (\ref{lagrang}) has the form of a Yang-Mills model for 
$\suinf \times G$ symmetry in the $\mathbb{R}^{(3+1)}$ curved parameter space - spacetime - of the 
subsystems:
\bea
\mathcal {L}_{U_s} & = & \int d^4x \biggl [\frac{1}{4} \tr (F^{\mu\nu} F_{\mu\nu}) + 
\frac{1}{4} \tr (G^{\mu\nu} G_{\mu\nu}) + \frac {\mathcal{M}}{2} \tr (\sD \rho_s) \biggr ], 
\quad \mu, \nu \in {0, 1, 2, 3} \label{yminvarsub} \\
F_{\mu\nu} & \equiv & F_{\mu\nu}^{lm} \hL^{lm} \equiv [D_\mu, D_\nu], \quad D_\mu = \partial_\mu - \Gamma_\mu - 
\sum_{lm} A_\mu^{lm} \hL^{lm}, \quad F_{\mu\nu}^{lm} F^{\mu\nu}_{lm} = L^*_{lm} L^{lm}. \label{yminvardefsuinf} \\
G_{\mu\nu} & \equiv & G_{\mu\nu}^a \hT^a \equiv [D'_\mu, D'_\nu], \quad D'_\mu = \partial_\mu - \Gamma_\mu - 
\sum_a B_\mu^a \hT^a, \quad G_{\mu\nu}^a G^{\mu\nu}_a = T^*_a T^a. \label{yminvardefg} 
\eea
The dimensionful constant $\mathcal{M}~ \propto ~M_P^n$, and similar to $\sD$ its value depends on the 
representation of Lorentz group of the parameter space - spacetime - realized by subsystems states 
$\rho_s$. The expression for $\sD$ would be similar to the examples given in Sec. \ref{sec:evolwhole} 
with an additional interaction term for $G$ symmetry. Equations (\ref{separation} - \ref{metric}) 
show how the metric of the parameter space is related to quantum states of the subsystem obtained from 
action (\ref{yminvardefg}).

\subsubsection{\bf Classical limit}  \label{sec:classlim}
When experiments are not sensitive to quantum field strength $F^{\mu\nu}$ of the $\suinf$ symmetry, only 
its effect on the geometry of the (3+1))D parameter space - the spacetime - described by 
(\ref{separation} - \ref{metric}) would be observable. Using (\ref{eulerint}), in Appendix (\ref{app}) 
we show that in classical limit the pure $\suinf$ term in (\ref{yminvarsub}) can be approximated by 
the 4D Ricci scalar $R^{(4)}$, which its integration over the 4D parameter space is no longer topological:
\be
\int d^4x ~\tr (F^{\mu\nu} F_{\mu\nu}) ~ \autorightarrow{classical}{limit} ~ {\large \propto} 
\int d^4x R^{(4)}  \label{r4}
\ee
This last step finalizes our demonstration that in \sqgr Einstein equation is a property of the 
parameter space characterizing the underneath quantum states of the Universe and its subsystems - matter. 
It confirms that Einstein equation should be considered as an equation of state~\cite{greos} and 
its quantization, as well as the quantization of spacetime are meaningless. Moreover, as the quantum 
gravity interaction has the form of a Yang-Mills model, its effect at high energies should resemble to 
additional gauge interactions on a curved spacetime. Therefore, it is also meaningless to talk about 
quantum corrections to the Einstein equation. In any case, it is well established that any change in 
Einstein equation can be considered as a change in geometry part or matter part, corresponding to Jordan 
or Einstein frame, respectively.

We should remark that in the above formulation it is assumed that multiple copies of the quantum clock 
are available for estimating the average value of an observable used to define the time parameter $t$. 
In other words the clock is tomographically complete. This is not a necessity, and time and/or relative 
distance may be quantified by non-projective measurements. We leave the investigation of such general 
case to future works. We do not discuss either the origin of dark energy/cosmological constant in this 
framework here, because it may depend on the quantum aspects of clock and reference and the fact that 
after their selection the Universe must be considered as an open quantum system.

A note is in order about the finding that in \sqgr quantum gravity is a Yang-Mills model. This means 
that its mediator quantum field is a vector - in the parameter space - rather than the observed spin-2 
graviton field of the classical Einstein gravity. Nonetheless, the relation (\ref{r4}) shows that there 
is no contradiction between the two observations. This is analogous to the predictions of the early 
models for strong interaction before the discovery of QCD model. Due to the strong coupling at low energies 
and confinement of constituent partons, the observations seemed to show a nonlocal and geometrically 
extended interaction analogous to a string. We now know that this phenomenological interpretation 
is wrong, and the confusion is caused by non-perturbative nature of the QCD interaction at energy scales 
lower than $\Lambda_{QCD}$. In the same manner, the deformation of spacetime, which in general relativity 
is interpreted as gravity, is generated by relative variation of quantum states of all constituents 
of the Universe, and the local metric and curvature of the parameter space - spacetime - present 
their {\it average} effect.

\subsection{\bf Summary of \sqgr model and its properties} \label{sec:sqgrresults}
We conclude this section by summarizing the \sqgr model and what is found about QGR in this framework so far:
\begin{itemize}
\item Assuming that Hilbert spaces of the Universe and its subsystems represent $\suinf$ symmetry, we 
showed that the Hilbert space of the Universe as a whole can be parameterized by 2 continuous parameter. 
When the Universe is divided to subsystems presenting a finite rank symmetry group $G$, and a quantum 
reference subsystem and a quantum clock are chosen, 2 additional parameters arise: a relative distant 
and a relative time \`a la Page \& Wootter or equivalent proposals. 
\item We interpreted the above 4D parameter space as the classical spacetime and demonstrated that 
its signature must be negative, i.e. it has a Lorentzian metric. Moreover, as it is a parameter space, 
its quantization is meaningless.
\item The coordinate independent affine parameter of the spacetime is related to the variation of the 
quantum state of the subsystems.
\item We defined symmetry invariant functionals over the Hilbert space of the Universe as a whole, 
and over those of its subsystems. They plays the role of an action for the evolution of the Universe and 
its subsystems, respectively.
\item The action has the form of Yang-Mills gauge theories on the parameter space for both $\suinf$ and 
subsystem specific {\it (internal)} finite rank $G$ symmetry. Thus, at quantum level like other forces 
mediator boson of gravity is spin-1.
\item We showed that the action functional for the whole Universe is static. Moreover, its purely 
$\suinf$ Yang-Mills part is topological and proportional to the Euler constant, i.e. integral over the 
2D Ricci scalar curvature. The constant of the proportionality is not an observable.
\item When the Universe is divided to subsystems, in the classical limit when the quantum Yang-Mills 
vector field of the $\suinf$ symmetry cannot be detected, the purely $\suinf$ Yang-Mills part of the 
action functional will be proportional to the 4D Ricci scalar curvature. Therefore, the classical limit 
of \sqgr is the Einstein gravity and the observed spin-2 graviton is a classical effective field.
\item This important prediction should be testable with future quantum experiments, for instance those 
seeking decoherence or entanglement initiated by quantum gravity.
\end{itemize}

\section{\bf Comparison with other quantum gravity proposals}  \label{sec:comp}
In this section we compare \sqgr with LQG and related models, string theories and related models, AdS/CFT 
conjecture, and several Quantum First models. This list is far from covering all the QGR proposals, so do 
the citations. In particular, non-commutative spacetime models, causal sets, and quantum gravity models 
based on the quantum histories are not compared with \sqgr, because of their fundamental differences. 
Nonetheless, some of them are briefly mentioned because of their connection with models reviewed here.

For each model we first remind its main assumptions and results, only for the purpose of fixing notations 
necessary for the comparison with \sqgr. We should emphasize that for the sake of briefness various new 
ideas and methods added to the original construction of these models are not explored here. Giving the 
fact that some of these proposals have been intensively under investigation for decades, their detailed 
review and comparison with \sqgr need a much extended work than this article. Moreover, our goal here is 
finding common features of the models rather than assessing their performance. For these reasons only 
the most foundational aspects and results of the models are considered and compared with those of \sqgr. 
We should also remind that \sqgr is a recent and under development proposal and its properties are not 
fully investigated. For this reason, its comparison with other QGR models is limited to what is known 
about it. Notably, its application to various QGR related phenomena is left to future works. 

As we discussed in the Introduction, due to the close relation between gravity and geometry of spacetime 
in the classical general relativity and Einstein gravity, finding a quantum model for gravitational 
interaction has been usually considered to be equivalent to quantization of spacetime as a physical 
entity. A notable difference between \sqgr and other QGR models is the absence of a quantized 
background or quantized spacetime. This unique feature becomes fundamental when one tries to compare this 
model with other QGR proposals. Indeed, a direct comparison cannot be made. Thus, the purpose of this 
work is to investigate whether there are comparable or analogous features in these models. For instance, 
$SU(2)$ group is present in the construction of many QGR models, including \sqgr. Our aim is to clarify 
the origin of these sort of similarities, and investigate whether they are superficial and unrelated, 
or reflect deep relation among models, despite their apparent differences. 

\subsection{\bf Background independent models} \label{sec:bgindep}
Following the failure of coordinate dependent canonical quantization of Einstein-Hilbert 
equation~\cite{qgrlagrangian,qgrcanonical,qgrcanonical0} (see e.g.~\cite{qgrearlyhist} for a review) 
and ADM (3+1)D description of Einstein equation and its quantization~\cite{admgr}, in 1961 Tullio Regge 
proposed a discrete but coordinate independent description of Einstein gravity~\cite{curvatureregge}. 
This model is the basis of most background independent QGR models. For this reason we briefly review 
it here. 

\subsubsection{\bf Regge discrete geometry}  \label{sec:regge}
According to this model a curved two or higher dimensional space can be approximately considered as flat 
everywhere except on the triangulated 2D surfaces - 2-simplexes. In particular, 3D or (2+1)D curved 
spaces can be approximated by sticking together tetrahedra with Euclidean or Lorentzian geometry in 
their bulk. The deficit angle of a vertex in the bulk of space is $\varepsilon = 2\pi - \sum_f \theta_f$ 
where $\theta_f$ is the angle of triangle (face) $f$ attached to vertex $v$, see e.g.~\cite{reggecal} 
for a review of Regge calculus. For vertices sitting on the boundary of the symplectic surface the 
deficit angle is $\varepsilon = \pi - \sum_f \theta_f$. The discretized gravity Regge action is: 
\be
S_{Regge} = \sum_e l_e \varepsilon_e \label{reggeaction} 
\ee
where index $e$ run over all edges, $l_e$ is length of the edge $e$, and $\varepsilon_e$ is the deficit 
angle of the vertex opposite to it. In Regge action tetrahedra edges can take any positive real value. 

\subsubsection{\bf Ponzano-Regge 3D QGR}  \label{sec:ponzanoregge}
In 1968 Ponzano and Regge proposed a 3D discretized quantum geometry model~\cite{qgrponzanoreggge} 
based on the Regge action $S_{Regge}$. They showed that if in (\ref{reggeaction}) $l_e$'s are chosen to 
be quantized spin, that is $l_e = j_e,~j_e \in \{0, 1/2, 1, 3/2, \cdots \}$ and $j_e$'s of each face 
satisfy triangle rule: 
\be
|j_1 - j_2| \leqslant j_3 \leqslant j_1 + j_2  \label{spintriangle}
\ee
their 6j symbol will be nonzero and approximately equal to the cosine of Regge action. 

Partition function of the Ponzano-Regge QGR is constructed from multiplication of the positive exponent 
of the cosine of Regge action for all tetrahedra, weighted, and summed over all configurations of spins:
\be
{\mathcal Z}_{PR} = \lim_{N \rightarrow \infty} \sum_{j \leqslant N} \Lambda^{N_0}(N) 
\prod_{e \in S_1} (-1)^{2j_e} (2j_e+1) \prod_{t \in S_3} (-1)^{-\sum\limits_{\hspace{5mm}e = 1, \cdots 6} j_e} 
\begin{Bmatrix}
j_1 & j_2 & j_3 \\
j_4 & j_5 & j_6 
\end{Bmatrix} \label{pgpartition}
\ee
Ponzano-Regge discrete quantum gravity was the first evidence of a close relation between gravity in 
3D space or (2+1)D spacetime and representations of $SU(2)$ group. This relation was later confirmed by 
the introduction of Ashtekar variables~\cite{ashtekarvar} in the framework of (3+1)D ADM formulation 
for quantization of gravity. In fact, as we explain in the following sections, the concept of 
triangulation and associating spins to edges of triangles comes up in one way or another in other 
QGR models, as well.

\subsubsection{\bf Ashtekar variable and Loop Quantum Gravity} \label{sec:ashtekarvar}
Loop Quantum Gravity (LQG)~\cite{lqgrev,lqgrev0} can be considered as continuum limit of symplectic 
QGR models~\cite{lqgrreggerev}. It uses ADM (3+1)D formalism with background-independent Ashtekar 
variables~\cite{ashtekarvar}. They consist of a spin connection 2-form $\omega_i^a (x)$, defined on the 
product of a 3D Euclidean manifold and a $SU(2)$ group manifold - more precisely a $SU(2)$ bundle on 
a 3D Euclidean manifold - and triads $E^i_a$ such that $E^i_a E^b_j= \delta_j^i \delta_a^b$ where 
$i = 1,~2,~3$ is coordinate index of the Euclidean space and $a = 1,~2,~3$ indicates generators of 
$SU(2)$ symmetry group. They replace coordinates and metric as dynamical variables. In the quantized 
model their dual variables are respectively $E^i_a$.and gauge field $A_i^a = \omega_i^a + \gamma K_i^a$, 
where $K_i^a \equiv K_{ij} E^{ja} /\sqrt{|h|}$, $K_{ij}$ is extrinsic curvature tensor of the 3D space, 
$h$ is determinant of the metric of physical 3D space, and $\gamma$ is the Immirzi 
constant~\cite{immirziparam}. 

\subsubsection{\bf $SU(2)$ symmetry, degeneracies and observables in LQG}  \label{sec:lqgsu2}
Although metric, and thereby coordinates, are apparently present in the definition of Ashtekar 
variables, their choice do not affect geometry of space and its quantization. The reason is that 
space curvature is described by $SO(3) \cong SU(2)$ transformation of a rigid frame, rather than 
deformation of the metric. Specifically, the rigid frame rotates when it is transported across 
the curved space manifold. On the other hand, the freedom of choice of the rigid frame at each point 
of the 3D manifold means that its $SO(3) \cong SU(2)$ symmetry is a gauge symmetry. Thus, $A_i^a$ 
and $E^i_a$ include more degrees of freedom than $g_{\mu\nu}$ in the (3+1)D classical gravity. This 
is evident from counting the number of components of these fields. 

To eliminate degeneracies observables of LQG and spin network 
(or foam)~\cite{qgrspinfoam,qgrspinfoam0,lqgfoam} - its discretized version - are quantized topological 
quantities generated by Wilson loops~\cite{lqgareaop}. This is why the model is called {\it Loop} QGR, 
and one of its most remarkable prediction is the quantization of area~\cite{lqgareaop}. This feature 
establishes the relation between LQG formulation using continuous Ashtekar variables, spin network as 
its approximation, and symplectic geometry of Ponzano-Regge: Quantized surfaces have non-trivial 
$SU(2)$ holonomy and triangulated 3D space \`a la Regge becomes a manageable approximation, including 
essential properties of a quantized curved space with meaningful continuum limit. 

\subsubsection{\bf Analogies between foundations of LQG and related models with \sqgr} \label{sec:lqganal}
In \sqgr conserved areas of diffeo-surfaces and their comparison induce an area (length) scale in 
the model, without being quantized. Moreover, $E^a_i$ fields are analogous to amplitudes $L_{l,m}$ in 
\sqgr. In fact, in~\cite{houriqmsymmeos} we show that in order to be invariant under coordinate 
transformations of the parameter space, these amplitudes must be differential operators in 
the parameter space. Indices $(l,m)$ are analogous to the {\it internal} $SU(2)$ symmetry of triads. 
However, in contrast to Ashtekar variables, their values are obtained from ensemble of representations 
of $SU(2)$ factors in the decomposition of $\suinf$ in equation (\ref{llmdef}). This property is 
similar to Ponzano-Regge and spin network where edges of tetrahedra are weighed by spins. However, in 
\sqgr both $l$, and $m$ quantum numbers of $SU(2)$ representations are involved in the action of the 
model and they are not constrained. The reason is that in contrast to LQG and Ponzano-Regge models, in 
\sqgr the Hilbert space does not represent a real space geometry.

\subsubsection{\bf Hilbert spaces of LQG and related models}  \label{sec:lqghilbert}
6j symbols consist of summation over weighted multiplication of 4 Wigner 3j symbols. In turn 3j symbols  
are proportional to Clebsch-Gordan coefficients $\langle j_1,m_1 ; j_2,m_2 | j_3,m_3 \rangle$, where 
$j_1,~j_2,~j_3$ respect triangle condition (\ref{spintriangle}). Therefore, each term in the partition 
function of Ponzano-Regge model (\ref{pgpartition}) is proportional to the projection of a $N$-spin to a 
one-spin state constrained by the triangle relation (\ref {spintriangle}) between {\it adjacent} spin 
states. 

Considering the expansion (\ref{llmdef}) of $\suinf$ group, it is clear that the Ponzano-Regge 
partition function ${\mathcal Z}_{PR}$ includes special configurations of a quantum system which its 
Hilbert space represents $\suinf$ symmetry, namely states that can be arranged as tetrahedra 
in a 3D space. This observation can be extended to 
other models based on a symplectic representation of space such as LQG, spin network, and Group Field 
Theories (GFT). Indeed, ~\cite{lqgtetrahed} describes explicit construction of the Hilbert space of a 
single tetrahedron in LQG/spin network by associating $SU(2)$ operators to edges of the tetrahedron. 
State of a unit cell of space - sometimes called {\it atom of space} - is generated by application of 
these operators to a vacuum state, such that the projection (amplitude) of the total spin of the 
tetrahedron is equal to its associated 6j symbol. This procedure can be extended to ensemble of 
$N \rightarrow \infty$ tetrahedra content of space, which can be also considered as spin-weighted 
graphs~\cite{lqghilbert}. Thus, we conclude that state generator operators, and thereby Hilbert spaces 
of Discrete QGR (DQGR) models such as Ponzano-Regge and LQG models, which we collectively call 
$\hm_{DQGR}$, are subspaces of the Hilbert space of a quantum system with $\suinf$ symmetry, such as \sqgr.

\subsubsection{\bf Kinematical and physical Hilbert spaces and reality conditions}  \label{sec:lqghilbert} 
It is useful to remind that 3j, 6j, and fundamental representation of $SU(2)$ are in general defined on 
the field of complex numbers. By contrast, a partition function or path integral over geometries of 
the physical space or spacetime, which should approaches to Einstein gravity in the limit of 
$\hbar \rightarrow 0$ must be real valued~\cite{lqghilbert,lqgfoamreality}. Moreover, due to the 
degeneracies discussed in Sec. \ref{sec:lqgsu2}, the Hilbert spaces $\hm_{DQGR}$ of LQG and related 
models are not {\it physical}, but {\it Kinematical}~\cite{lqghilbert}. The Hilbert space of physical 
states $\hm_{phys}$ containing quantized background independent geometries is a subspace of $\hm_{DQGR}$, 
that is $\hm_{DQGR} \supset \hm_{phys}$. However, it is in general difficult to construct $\hm_{phys}$ 
explicitly~\cite{lqghilbert}. In addition, demonstration of diffeomorphism and Lorentz invariance of 
physical states is not straightforward and one might expect violation of Lorentz invariance in QGR 
models with discretized space~\cite{lqglorentzviol}. Indeed, diffeomorphism invariance of DQGR is 
explicitly shown only for special cases~\cite{lggcovar,lggcovar0}. 

Even in DQGR/LQG models that preserve Lorentz invariance dispersion relation of gravitational 
waves~\cite{lqggwdispersion} and electromagnetic radiation~\cite{lqgphenomnrev} may deviate from 
general relativity. However, both of these deviations are stringently 
constrained~\cite{grestgrb090510a,gwdisperligo,gwdisperligo0}. Moreover, Immirzi parameter may affect 
interaction of fermions~\cite{lqgimmirzfermion}, and thereby induces a {\it fifth force} type effect 
on matter. This effect is also constrained by various tests of gravity~\cite{grfifthconstraint}.

Complexities analogous to nonphysical states in the formulation of LQG and related models do 
not arise in \sqgr. The parameters defining the Hilbert space, namely $(t, r, \theta, \phi)$ are real, 
and by construction their redefinition - in other words diffeomorphism of the parameter space - 
corresponds to change of the Hilbert space's basis by application of a unitary transformation - a 
member of $\suinf$ symmetry group of the subdivided quantum Universe. We also notice that although \sqgr, 
Ponzano-Regge model, LQG, spin network, and their extension to GFT share $SU(2)$ symmetry in their 
construction, in practice all of them, except \sqgr, use only the Casimir operator of $SU(2)$. 
The reason is that eigen states $m$ of azimuthal projection of spin vector induce a preferred 
direction or in other words a frame, which these models want to avoid.

\subsubsection{\bf Time and matter in LQG} \label{sec:lqgtime}
Similar to \sqgr, in LQG and related models time must be considered as a relational observable. 
One way of making the model dynamic is to consider time as the classical affine parameter of 
histories~\cite{lqghistories,lqghistories0} or path integrals in the quantized physical 
space~\cite{lqgtime}. Although in such setups Lorentz and diffeomorphism invariance is not trivial, it 
may be achievable~\cite{lqgtime,lggcovar,lggcovar0}. 

Describing time by histories needs a {\it historian} - a reference subsystem with respect to 
which histories are defined. But, construction of background independent QGR models do not clarify how 
to satisfy necessary conditions for division of a quantum system~\cite{sysdiv}. In fact kinematical 
Hilbert space $H_{DQGR}$ seems to be inseparable~\cite{lqghilbert}. Specifically, division of the Hilbert 
space to orthogonal blocks, which could be considered as subsystems, needs an  additional symmetry, 
because in these models $SU(2)$ is inherently related to gravity. We might consider tetrahedra as 
the most fundamental {\it atomic} subsystem~\cite{lqgcoarse}. However, to discriminate one tetrahedron 
as reference, there must be selection criterion, thus another symmetry - observable. This issue is 
directly related to the fact that LQG and related models do not consider matter fields - a symmetry 
orthogonal to space - in their foundations. Although, a time parameter and matter fields can be easily 
added to the Einstein gravity Lagrangian described as a function of Ashtekar variables and their duals, 
see e.g..~\cite {lqgmatter}, the foundational issue of time definition in LQG and related models is 
not fully solved. Attempts to solve this problem, for instance through quantization of phase 
space~\cite{lqgphase,lqgphase0,lqgphase1}, indeed include matter and/or symmetries orthogonal to 
diffeomorphism symmetry.

\subsubsection{\bf Non-perturbative characteristic of LQG and related models}  \label{sec:lqgnonperturb}
The origin of subsystem definition issue in background independent models is their non-perturbative 
approach to QGR. Division to subsystems needs a criteria for breaking the Hilbert space or its parameter 
space to distinguishable sectors. Such operation implies the possibility of a perturbative description 
of the system at some scale. However, in LQG and related models, in absence of matter there is no natural 
covariant rule for a quantum gravitational perturbative expansion. This observation clarifies why 
there is no inherent way to include matter in these models. In fact, division to subsystems; emergence 
of a quantum clock, inclusion of matter in the foundation of the model, and existence of both perturbative 
and non-perturbative regimes are related. In \sqgr they are naturally implemented in the construction of 
the model through special form of its symmetries.

\subsubsection {\bf Outline of comparison between background independent models and \sqgr} \label{sec:lqgsum}
In conclusion, although $SU(2)$ symmetry plays an essential role in the construction of background 
independent models and \sqgr, its role and {\it raisons d'\^etre} in these models are very different. 
Notably, in LQG, GFT, and other symplectic models it is strictly related to the assumption of a physical 
3D quantum space. Nonetheless, spin network realization of LQG can be considered as a subspace of the 
Hilbert space of \sqgr, in which with additional relations - entanglement - between representations of 
$SU(2)$ components are considered. Both background independent models and \sqgr rely on the definition 
of a relative time or histories, which need division of the Universe to subsystem. In \sqgr this 
concept is built in the construction of the model and provides the necessary ingredients for definition 
of a quantum subsystem as clock and inclusion of matter fields.

\subsection{\bf Quantum approaches to QGR}  \label{sec:qmqgr}
Inherently quantum approaches - called {\it Quantum First} by some authors~\cite{qgrlocalqm} - are 
relatively recent arrivals into the jungle of QGR proposals and \sqgr can be classified in this group. 
For this reason it is crucial to investigate its similarities and differences with other models in this 
category. 

A shared characteristic of Quantum First models is the absence of a classical spacetime as a foundational 
concept in their axioms - or at least this is the claim. Consequently, it has to emerge down the road 
from more primary properties and structures of an abstract quantum system. It is useful to remind that 
the concept of an emergent spacetime is not limited to these models. The possibility that spacetime 
may not be a fundamental entity is also considered by other QGR candidates as well, see 
e.g.~\cite{spaceemerge,spaceemerge0,spaceemerge1}. Specifically, it is suggested that a quantum Lorentz 
invariant spacetime orthogonal to internal gauge symmetries may emerge in 
QGR models based on the extension of the Poincar\'e group and gauge 
symmetries~\cite{qgrgaugesep,qgrgaugesep0,qgrgaugesep1}. The idea of spacetime emergence is also 
explored by models in which, in one way or another, thermodynamics and quantum gravity are 
unified~\cite{qgrthermo,qgrthermo0,qgrthermo1}. These models seem to have little common aspects with 
\sqgr and we do not discuss them further here. 

In absence of any hint about the quantum nature of gravity, for instance its Hilbert space, and its 
relationship with classical gravity and other interactions, Quantum First models usually use priors 
inspired from semi-classical gravity, in particular from properties of semi-classical physics of black 
holes. Based on these priors two categories of Quantum First models other than \sqgr can be 
distinguished: 
\begin{description}
\item {-} Models that consider locality and causality as indispensable for QGR: Some of these models 
need modification of standard quantum mechanics;
\item {-} Models inspired by black hole entropy and its relationship with holographic principle and 
AdS/CFT duality.
\end{description}

\subsubsection{\bf Modified quantum mechanics and locality} \label{sec:modqm}
{\it Locality} is considered to be crucial for describing black holes, their 
thermodynamics~\cite{bhentropy,hologprin} and its puzzles~\cite{hawkingrad,bhentropyloss}. More 
generally, causality and observed finite speed of information propagation in both classical general 
relativity and QFT implies some degree of locality in any interaction, including QGR. For these reasons, 
locality and its close relationship with the definition of subsystems as {\it localized} entities in the 
Universe have been the motivation of authors of~\cite{qgrhistory,qgrhistory0,qgrlocalqm} for proposing a 
{\it generalized} quantum mechanics. Specifically, history description of quantum 
mechanics~\cite{qmhistories,qmhistories0} is generalized in~\cite{qgrhistory} to define 
{\it coarse-grained} histories as a bundle of {\it fine-grained} histories (path integrals). They replace 
the Hilbert space of quantum mechanics, which in a QGR framework corresponds to a spacelike surface 
during an infinitesimal time interval, defined with respect to a reference clock. In addition, in this 
modified quantum mechanics projection operators to eigen states of position are time-dependent, and 
during each time interval they project states to a different set of histories. In turn, sets of histories 
present subspaces of the bundle. Presumably, in this model not only the state of a system, but also its 
whole Hilbert space changes with time. 

Inspired by generalized quantum mechanics,~\cite{qgrhistory0} proposes an alternative way to implement 
locality in what is called {\it universal} quantum mechanics. In analogy with the bundle space 
of~\cite{qgrhistory} it extends the space of physical states to provide additional labeling, such 
as {\it in and out} states in curved spacetimes~\cite{qmcurve}. In addition, labels can be interpreted 
as time or labels of states in a multiverse, as needed. Physical states can be considered as 
{\it local} in this extended state space. 

These models and other QGR proposals based on the quantum histories, 
see e.g~\cite{qgrhistoryrev,qgrhistorymultivers} and references therein, have little common features 
with \sqgr, which is strictly based on the highly tested standard quantum mechanics. The reason for having 
reviewed them here is their role in the development of further models with some similarities with \sqgr, 
which we will review in the following subsections.

\subsubsection{\bf QGR from locality and causality} \label{sec:qgrlocality}
Localization of quantum mechanics in~\cite{qgrhistory} does not specify an explicit implementation 
procedure. Nonetheless, motivated by this model~\cite{qgrlocalqm,qgrlocalqm0,qgrlocalqm1,qgrlocalqm2} 
propose a road-map for realization of this concept in what they call Local Quantum Field Theories (LQFT).  
In these QFT models observables convoy quantum information only locally. Here we call the corresponding 
QGR proposal LQFT-QGR.

In quantum systems with infinite degrees of freedom, such as in QFTs, spacetime sector of the Hilbert 
space cannot be factorized to disconnected (untangled) subspaces without violating causality. Such 
quantum systems are said to have Type III operator algebra in the classification 
of~\cite{vonNeumannoptype,type3op}. For this reason, in LQFT-QGR the division to subsystems is performed 
algebraically. Specifically, it is assumed that for any region of spacetime $U$ there is an extension 
$U_e$. Observables $\hA$ and $\bar{\hA}$ are defined such that they have nonzero support respectively 
on $U$ and $\bar{U}_e$, where $\bar{U}_e$ is the complementary space of $U_e$. Under these conditions 
$\hA$ and $\bar{\hA}$ are assumed to be disentangled in a specific vacuum:
\be
\langle U_e | \hA \bar{\hA} |U_e \rangle = \langle 0 |\hA|0 \rangle \langle 0 |\bar{\hA}|0 \rangle 
\label{giddingsaa}
\ee
The vacua $|U_e\rangle$ and $|0\rangle$ are related by a Bogoliubov transformation. This definition is 
considered to provide a sort of localization without factorization of the Hilbert space. However, it 
is evident that this algebraic structure is not in general diffeomorphism invariant and observable 
operators $\{\hA\}$ and $\{\bar{\hA}\}$ must satisfy specific conditions to retain their invariance and 
physical meaning~\cite{qgrlocalqm2}. Seeking such operators,~\cite{qgrlocalqm0,qgrlocalqm2} find that in 
analogy with gauge invariant Wilson loops in Yang-Mills theories, diffeomorphism invariant operators 
$\Phi_\Gamma \in \{\hA\}$ are nonlocal structures, which depend only on the spacetime 
connection~\cite{qgrlocalqm0}. Specifically:
\be
\Phi_\Gamma (x) = \phi (x^\mu + V^\mu_\Gamma)  \label{giddingsscalar}
\ee
where $\Gamma$ is a path running from point $x$ of the spacetime to infinity, and $V^\mu_\Gamma$ is the 
integral of an expression depending on the metric of spacetime along the path $\Gamma$. An explicit 
expression for $V^\mu_\Gamma$ is obtained for the weak coupling limit of semi-classical gravity 
in~\cite{qgrlocalqm2}.

\subsubsection{\bf Comparison of LQFT-QGR with \sqgr}  \label{sec:lqftcomp}
In LQFT-QGR two essential concepts for QGR, namely division of the Universe to subsystems and carriers 
of quantum information are considered to be the same. In this respect, the model is similar to \sqgr, 
that is carriers of information are matter/radiation fields and their internal symmetries, which is 
orthogonal to diffeomorphism of spacetime and a necessary criteria for division of the Universe to 
subsystems. However, the two models are conceptually very difference. In LQFT-QGR subsystems are 
somehow localized in spacetime. By contrast, in \sqgr spacetime is not the quantum Universe and no 
locality condition is imposed on subsystems/particles. In fact, in \sqgr locality and causality are 
not postulated. As we discussed in Sec. \ref{sec:geom}, they arise from quantum uncertainties. 
Moreover, interpretation of coordinates in (\ref{metric}) as average or expectation values, shows that 
in agreement with quantum mechanics observations, locality in general is an approximation.

LQFT-QGR and \sqgr share the absence of a classical dynamics in their foundation. Moreover, 
both models are a type of QFT on a curved spacetime, which plays the role of a parameter space. Their 
difference is in the definition of observable fields: LQFT-QGR constrains field operators to realize 
special algebraic structures and a sort of locality, whereas in \sqgr both gravity and matter sectors 
are quantum fields similar to QFTs without gravity. In addition, in \sqgr spacetime genuinely emerges, 
whereas in LQFT-QGR it is implicitly postulated and present from the beginning. Although in contrast to 
many other QGR proposals spacetime per se is not quantized. the model offers no explanation for its 
origin, its dimension, or properties of its metric, or its relationship with other quantum fields.

\paragraph*{\bf Type III algebra in LQFT-QGR and \sqgr}
Operators indexed or parameterized by $\mathbb{R}^n$ cannot be divided to subsets associated to limited 
regions of the indices, if the whole algebra has to be invariant under 
diffeomorphism~\cite{vonNeumannoptype,type3op}. It is why a symmetry orthogonal to diffeomorphism is 
necessary for {\it tagging} and fulfilling conditions for definition of quantum subsystem~\cite{sysdiv}. 

As QFTs, both LQFT-QGR and \sqgr are Type III quantum systems. In \sqgr the inseparability of continuous 
operators is reflected in the common $\suinf$ symmetry of all subsystems, including the Universe as a 
whole, and the need for a factorized finite rank {\it internal} symmetry. By contrast LQFT-QGR considers 
strict locality as a foundational concept and tries to use nontrivial topological structures as a 
replacement for {\it tagging} and identifying subsystems. However, at present there is no evidence for 
the possibility of such algebraic structures in QFTs, except for the solutions obtained in the weak 
coupling regime of semi-classical gravity~\cite{qgrlocalqm2}. Moreover, although topological structures 
are observed in condensed matter, they are extremely fragile. By contrast, symmetry breaking or 
emergence, as requested in \sqgr, is wide spread in nature. We also notice that topological structures 
proposed by LQFT-QGR are different from those used in LQG as observables. In LQG Wilson loops do exist 
because of axioms and construction of the model. By contrast, the existence of such operators LQFT-QGR 
are in large extend a conjecture. The model explored in~\cite{qgrlocalqm2} for such structures is 
semi-classical and includes perturbative Einstein equation, which is non-renormalizable and cannot be 
considered as a genuine QGR.

\subsubsection{\bf QGR and emergent spacetime from entropy and holography} \label{sec:qgrentropy}
Another set of conjectures used for getting insight into QGR without considering an underlying classical 
dynamics is the holographic principle~\cite{hologprin,hologprin0,hologprin1} and gauge-gravity duality 
conjecture~\cite{qgrgaugedual,qgrgaugedual0}, specially in the form of AdS/CFT duality, see 
Sec. \ref{sec:adscft} for more details. Notice that this conjecture should not be confused with models 
that try to quantize gravity by extending gauge group of the Standard Model, such that it includes 
Lorentz and Poincar\'e symmetries~\cite{qgrgaugesep,qgrgaugesep0,qgrgaugesep1}. 

Motivation for the holography conjecture~\cite{hologprin,hologprin0,hologprin1} is the proportionality 
of semi-classical black hole entropy to area of its horizon, rather than to its 
volume~\cite{bhentropy,hawkingrad}. According to holography conjecture there is an upper limit on the 
amount of quantum information contained inside the bulk of a region of spacetime~\cite{hologprin}. It 
is proportional to the area of its boundary and is maximal for black holes~\cite{bhentropy,hawkingrad}. 
This conjecture is not limited to gravitational systems and similar behaviour is observed in other 
many-body quantum systems, if a suitable null (light-like) boundary surface can be defined~\cite{hologrev}. 
In particular, entanglement entropy of some low dimensional many-body quantum systems at critical point, 
that is when the system is scale invariant and behaves conformally, is calculable analytically, and the 
results show that they follow holographic
principle~\cite{qgrautomatom,condmatterentropy,condmatterentropy0}.

AdS/CFT duality conjecture~\cite{stringgauge,adscftentangle} posits that quantum properties of the 
boundary of a spacetime region in the limit that it can be approximated by a conformal QFT can be 
related to geometry and QGR of the bulk if its background geometry is AdS.

Inspired by these conjectures,~\cite{qgrentangle} considers two quantum system with a quantum CFT living 
on their common boundary. An analogy is established between the reduction of entanglement entropy and 
exchanged quantum information between the two systems when their boundary is shrunk, and the reduction 
of gravitational interaction with increasing distance. To understand this analogy, imagine squeezing a 
rubber bar in the middle. More it is squeezed, more material is pushed to the two ends and smaller 
becomes the surface connecting them until the bridge breaks and the two parts separate. Of course, this 
analogy is very far from being a QGR model. Nonetheless, it has motivated construction of QGR models 
using entanglement entropy as the origin of what is classically perceived as geometrical distance. 

\subsubsection{\bf Entanglement-Based Models (EBM) of quantum gravity}  \label{secebm}
A more systematic approach to construction of a spacetime from entropy-area law is proposed 
in~\cite{qgrentangle1,qgrentangle2}, where spacetime metric and geometry emerge from tensor decomposition 
of the Hilbert space of the Universe to entangled subspaces. This model is based on several axioms, 
see~\cite{qgrentangle2} for the complete list. They include: 

\begin{enumerate}
\item A preferred tensor decomposition of the Hilbert space $\hm$ [of the Universe], where each factor 
$\hm_i$ presents Hilbert space of a point or a small space around a point of space;
\be
\hm = \bigotimes_i \hm_i  \label{hilbertdecomp}
\ee \label{ebm1}
\item There is what is called Redundancy Constrained (RC) states for each subset of the Hilbert 
space $B \subset \hm$, considered to be a subspace of physical space. Its entropy is assumed to be:
\bea
S(B) & \equiv & \frac{1}{2} \sum_{i \in B, j \in \bar{B}} I(i:j) \label{ebmentropy} \\
I(i:j) & \equiv & S(i) + S(j) - S(i \cup j) \label{mutualinfo}
\eea
where $I(i:j)$ is the mutual information of subsystems $i$ and $j$. This construction replaces area-law 
axiom considered in~\cite{qgrentangle,qgrentangle1}. \label{ebm2}
\item It is assumed that the system is in an {\it entanglement equilibrium} state, when subsystems are 
in RC states. Under small perturbations the entropy of $B$ is assumed to be conserved. This means that 
the total entropy is conserved. Moreover, when states deviate from RC, their entropy can be decomposed 
to entropy of a fiducial RC state and a subleading component, interpreted as an effective field theory. 
The two components cancel each other to preserve the total entropy.  \label{ebm3}
\end{enumerate}
It is clear that axiom \ref{ebm1} is constructed such that the Hilbert space $\hm$ presents physical 
space. Thus, we conclude that similar to LQFT-QGR in this model the space does not really emerge, but 
its existence is postulated. Moreover, we notice that the definition of subsystems is loose and does 
not explicitly respect necessary conditions~\cite{sysdiv}. It is why this axiom explicitly states that 
factorization is static and somehow is {\it preferred}. But it is not specified what is the criteria 
for its selection.

Axiom \ref{ebm3} replaces action and variation principle that in classical mechanics and QFT models 
lead to dynamics and field equations, respectively. In addition, according to this axiom RC states can 
be considered as a {\it background} around which a perturbation is performed. Indeed, the model does 
not consider highly non-RC states and applies only to weak gravity cases~\cite{qgrentangle2}.

The structure described by above axioms can be considered as an information graph, which its vertices 
are factors of the Hilbert space and its edges are weighted by mutual information $I(i:j)$ of subsystems 
corresponding to factors of the Hilbert space. This graph is analogous to discrete geometry in 
Ponzano-Regge, spin network, and LQG.  

To complete the geometrical interpretation, the area of information graph or its subgraphs must be 
related to entanglement information. In~\cite{qgrentangle,qgrentangle1} this connection is established 
by assuming holographic principle. However, when RC structure is assumed~\cite{qgrentangle2}, according 
to one of the axioms of the model (axiom 3 in~\cite{qgrentangle2}), the area associated to a subspace 
$B$ of the space is:
\be
\mathcal{A} (B, \bar{B}) = \frac{G_N}{2} I(B:\bar{B})    \label{ebmarea}
\ee
where $G_N$ is the Newton constant (for $\hbar = 1$ and $c = 1$) and $\bar{B}$ is the complementary of 
$B$. Although the area $\mathcal{A}$ associated to a subspace of the Hilbert space is not the boundary 
of a bulk space, the inspiration from holographic principle is evident. This axiom and Radon transform 
is used to describe area as a function of the entropy of factors $\hH_i \forall i$ of the Hilbert space 
and define a background metric. Perturbation of this metric are interpreted as the perturbation of 
quantum state of the physical space. 

Additionally, variation of the entanglement graph geometry is used as a {\it clock} to which a 
Hamiltonian and an operator analogous to energy-momentum can be associated. The latter can be considered  
as an effective field theory generating subleading entropy of states, which are perturbatively deviated 
from RC states. Finally, by comparing this formulation with general relativity and using Radon 
transform,~\cite{qgrentangle2} argues that Einstein equation can be concluded.

\subsubsection{\bf Comparison of EBM with \sqgr} \label{sec:ebmcomp}
We find that EBM is more similar to \sqgr - in spirit rather than construction - than other models. 
Here we briefly highlight their common features. 

\paragraph*{\bf Factorization of the Hilbert space and division to subsystems}
The importance of division of the Hilbert space to factors presenting subsystems is crucial in both 
models. However, as remarked earlier, in EBM the division is considered to be rigid and {\it preferred}. 
This is in strict opposition to the approach of \sqgr. The reason behind the special factorization is 
again the absence of a concrete criteria to discriminate between factors - subsystems. 

We notice that the issue of how to divide the Universe and its Hilbert space to quantum subsystems 
generally arises in quantum approach to QGR due to foundational requirements~\cite{houriqmsymm}, and in 
some other QGR models for various reasons. Model makers use different schemes to deal with this 
crucial matter. For instance, they introduce topological structures - as in LQG and LQFT-QGR; or simply 
consider a fixed decomposition without addressing its origin, as in EBM. \sqgr assumes an orthogonal 
finite rank symmetry - presumably from symmetry breaking or emerging - to fulfill general conditions for 
division of a quantum system to subsystems according to the criteria defined by~\cite{sysdiv}. Although 
the nature and origin of this symmetry is not specified in the construction of \sqgr, properties of 
$\suinf$ symmetry, notably equations (\ref {sundecompos}, \ref{suinfn}) facilitate the interpretation of 
the Universe as a many-body quantum system, in which based on our knowledge from condensed matter, a 
symmetry of the form (\ref{sqgrsymm}) can arise relatively easily. More importantly, in \sqgr the finite 
rank symmetry is associated to matter. In this way, matter and space become intertwined and 
inseparable. This is not the case in EBM, LQFT-QGR or LQG and related models. 

\paragraph*{\bf Geometry and classical gravity}
Another common aspect between EBM and \sqgr is the explicit dependence of the space geometry on the 
quantum state - through entanglement entropy in EBM and through fidelity in \sqgr. However, emergence, 
construction, and physical meaning of the space in the two models are very different. In EBM 
of~\cite{qgrentangle1,qgrentangle2} factors of the Hilbert space are considered to present points or 
regions of the physical space and the information graph is interpreted as a symplectic geometry, 
which in the continuum limit can be considered as a quantized space. Therefore, although the existence 
of a physical space is not explicitly mentioned in the axioms, it is implicitly behind the factorization 
of the Hilbert space. By contrast, in \sqgr space genuinely emerges as parameter space of $\suinf$ 
representations. 

A consequence of these differences is that \sqgr has an explicit explanation for the dimension of 
spacetime, where as in EBM dimension of the space(time) is not fixed. In fact, the information graph can 
be embedded in any space with dimension $d \geqslant 2$. Notice that the relation between area of a 
subgraph (subsystem) and its entanglement entropy with its complementary in (\ref{ebmarea}) does not 
restrict the graph to be planar - not even locally. A priori every vertex - that is every factorized 
subsystem of the Hilbert space - can be entangled to all other subsystems. In~\cite{qgrentangle1} 
it is assumed that the number of entangled subsystems to a vertex - corresponding to the number of edges 
attached to it - is limited. Nonetheless, their number can be large and the graph rules do not constrain 
their mutual angle. Thus, in contrast to Ponzano-Regge and LQG, in which spins associated to edges of 
the symplectic space must satisfy triangle constraint at each vertex, the information graph in EBM can 
be embedded in a multi-dimensional space. For these reasons, $d$ is considered as a stochastic parameter 
determined from averaging over geometries of many information graphs~\cite{qgrentangle1}. On the other 
hand, spacetime dimension is a fundamental quantity which affects many observables in particle physics 
and cosmology at all energy scales. So far no evidence of an extra/infra or stochastic dimension is 
detected.

In \sqgr the relationship between affine parameter, metric, and quantum fidelity in equation 
(\ref{separation}) naturally relates ensemble of parameters (not just distance or area) to quantum states 
of the subsystems. In both EBM and \sqgr Einstein equation remains classical and is obtained from 
relationship between quantities with underlying quantum origin.

\paragraph*{\bf Analogy between distance and entanglement}
In both models an area quantity emerges and it has a crucial role for their interpretation as QGR. 
In \sqgr it emerges from comparison of the preserved areas of diffeo-surfaces of subsystems with 
an arbitrary reference subsystem. In EBM it is postulated in (\ref{ebmarea}), where a dimensionful 
area/distance parameter is mandatory. Although, the way a scale emerges in these models is very different,  
in both cases it is related to the division of Universe to subsystems. Indeed, in EBM entanglement and 
its associated entropy are meaningful only when multiple quantum systems are present. In \sqgr division 
to subsystems is necessary to make the conserved area of diffeo-surfaces relevant and measurable.

In addition to difference in the manner that a dimensionful scale arises in these models, there is another 
important difference. In \sqgr the area is related to geometry of the compact parameter 
space of representations of $\suinf$ symmetry of subsystems. Thus, it is a well defined and unique 
measurable for each subsystem relative to a reference. By contrast, quantification of entanglement and 
relative quantum information is not unique and various definitions, e.g. von Neumann or R\'enyi entropy 
can be used, and each of them has its own merit and applications. EBM models  
of~\cite{qgrentangle,qgrentangle1,qgrentangle2} do not specify which one of these entropies should be used 
or what is rationale for preferring one to others, or whether different definitions should be interpreted 
as different choices of coordinates. 

\subsection{\bf String theory, M-theory, and AdS/CFT duality in 3 and higher dimensions} \label{sec:stringadscft}
String theory and related models are without doubt the most intensively studied QGR proposals. Although 
some of Quantum First models are inspired by (super)string theories and AdS/CFT duality conjecture, string 
theories are not properly speaking Quantum First. Their perturbative formulation is a quantized 2D sigma 
model, originally proposed for describing strong nuclear interaction~\cite{stringrev}. Non-perturbative 
formulation of string models, also called M-theory, and its realization as matrix model has the form of 
a (super)Yang-Mills QFT. 

In the recent decades new approaches to string theories are extensively studied in the literature and 
various concepts and structures are added to their initial construction. Their list includes: 
D-branes states~\cite{stringbranerev}; String condensation and its relation with p- and D-brane 
solutions~\cite{stringbranerev,stringcond,stringcond0,stringcond1}, important for the conjectured 
non-perturbative formulation of string models also called the M-theory; And AdS/CFT 
duality~\cite{stringgauge,adscftrev}, which is the simplest case of gauge-gravity 
conjecture~\cite{gaugestringcorr,stringgauge0,stringgauge1} and closely related to M-theory and matrix 
models. Nonetheless, the basic structure of (super)string theories and their properties continue to be 
considered as foundational and established knowledge for development of these more advanced theories. 
In particular, M-theory uses the 10D or (9+1)D Euclidean or Minkowski spacetimes, respectively. This is 
the fundamental dimension of spacetime in perturbative superstring theories. Similarly, the first 
evidence of AdS/CFT correspondence was discovered for D3 brane models in a 10D compactified 
$AdS_5 \times S_5$ background spacetime~\cite{adscft5d}. Thus, due to the importance of perturbative 
formulation of string models, in this section we first briefly remind their findings and how they compare 
with \sqgr. Then, we review and compare M-theory and its matrix realization, and AdS/CFT conjecture.

\subsubsection{\bf Perturbative string theories and their comparison with \sqgr} \label{sec:string}
As extended literature and textbooks on string theory related subjects such as~\cite{stringrev,stringrev0} 
are available, we do not review these models in details and only remind their most important properties 
used for comparison with \sqgr wherever they are necessary.

\paragraph*{\bf 2D surfaces in string theory and \sqgr}
Overlooking all the complexities of string and superstring theories, they can be summarized as 2D quantum 
gravity of a conformal quantum sigma model. Here quantum gravity means summation over all possible 
geometries of their 2D worldsheet - more generally a membrane. In this view of string theories, their 
most evident common feature with \sqgr is the crucial role of 2D surfaces and their diffeomorphism in 
their construction. As we will see in more details in Sec. \ref{sec:mtheorymatrix}, even in 
non-perturbative approaches such as M-theory and p- and D-brane models, the 2D surfaces do not lose their 
crucial role and are implicitly present and represented by large matrices. 

On the other hand, the role, properties, and interpretation of 2D surfaces in these theories and \sqgr 
are profoundly different. In string theories 2D worldsheets of strings or more generally membranes are 
quantized and summation over their geometries is interpreted as path integral of 2D quantum gravity. 
By contrast, diffeo-surfaces in \sqgr are not an independent physical entities, neither they are quantized. 
They are associated to quantum states of the Universe and its content - subsystems in the same way that 
in QFT a charge or spin is associated to a particle but it is not the particle. Deformations of 
diffeo-surfaces do not correspond to different (quantum)-gravitational states, but rather represent 
members of the symmetry group of shared by all quantum subsystems, including the whole Universe. 

\paragraph*{\bf String sigma model}
As a sigma model both bosonic and fermionic quantum fields live on the 2D worldsheet of 
strings~\cite{stringrev,stringrev0}. In superstring models they are interpreted as coordinates of a 
$n-$dimensional quantum spacetime and their supersymmetric counterparts, respectively. One can equally 
interpret the worldsheet of a string as a 2D extended membrane embedded or emerged in an $n-$dimensional 
spacetime. Additionally, string theories are in general 2D Conformal Field Theories (CFT). This means that 
they are invariant under rescaling of both worldsheet 2D coordinates and local rescaling of the fields. This 
double conformality is a necessary condition for eliminating central charge and anomalies, which arise 
when these models are quantized~\cite{stringrev0}. Cancellation of these unwanted elements limits the 
spacetime (target space) dimension to $n = 26$ for bosonic strings or to $n = 10$ in superstring models. 
More generally, the sigma model can be any CFT with Kac-Moody algebra having the same number of degrees 
of freedom as bosonic and supersymmetric models. As mentioned earlier, the value of fundamental - rather 
than observed - spacetime dimension obtained from sigma model formulation of superstrings is taken 
for granted in further developments of these models. Interestingly, $n = 1$ model is also a consistent 
quantum model~\cite{stringrev}. The single (super)field in such model cannot be interpreted as a 
background spacetime, but they are studied as a decoupled sector in matrix formulation of string 
theory~\cite{qgrmatrixu1}.

In the framework of \sqgr, the string setup - without quantization of 2D worldsheet/membrane - can be 
considered as a special state for a quantum system with $\suinf$ symmetry. The sigma model of strings - 
without constraints arising from conformal symmetry and quantization - can be interpreted as special 
states for subsystems with an internal symmetry $G$. Quantization of such a state in the framework 
(super)string theory restricts the internal symmetry $G$ to groups allowed by the cancellation of 
anomalies. For instance, $G$ may be identified with: 10D sigma model of superstring models and/or 
its $SO(32)$ or $E8$ internal symmetry; symmetries of the low energy corresponding $\nm = 4$ 
supergravity in 11D; or symmetries of compactified coordinates or quantum fluctuations of D-brane 
solutions in M-theory. The origin of these similarities can be traced back to the Virasoro algebra of 
string fluctuations, which is a subalgebra of Surface-preserving Diffeomorphism (SDiff) of a torus 
$SDiff (T^2)$ and the fact that the latter is a representation of $\suinf$ 
group~\cite{virasorosuinf,virasorosuinf0,suninftorus}. 

One of the main advantages of string theory to canonical QGR is its renormalizabilty and absence of UV 
singularity, owed to the extended nature of strings. Although details are not yet worked out for \sqgr, 
from its Yang-Mills action we expect that it be renormalizable. Moreover, UV singularity should not 
arise, because the distance between subsystems is related to the relative area of their diffeo-surfaces, 
which by definition cannot shrink to a point - equivalent to zero distance, other $\suinf$ would be 
represented trivially. This feature should play the role of a build-in ultra-violet cut-off without 
introducing any fixed scale.

\paragraph*{\bf Curved spacetime and gravity in string theory}
The issue of a curved target (field) spacetime in string theory does not have an analogy in \sqgr. 
However, solutions proposed to overcome this problem and their role in non-perturbative formulation of 
string theory in the framework of M-theory can be compared with \sqgr. 

In sigma model formulation of string theories quantization is consistent only when the geometry of the 
field (target) space - interpreted as fundamental spacetime - is flat. This means that only metric 
perturbations around this Minkowski background are quantum mechanically meaningful. Although in 
the framework of perturbative string theories a string {\it gas} has been considered - specially for 
the purpose of describing cosmological perturbations~\cite{stringgas} - the inherently intertwined 
nature of spacetime and strings may make it impossible to consider them as separately evolving entities. 
There are, nonetheless, exceptions. AdS/CFT duality conjecture, discussed in more details in 
Sec. \ref{sec:adscft}, is proved for AdS$_3$ space, and is considered as evidence for consistent 
formulation of string theory, at least in some curved background spaces.

Another way to overcome the issue of curved target space is considering special configurations/solutions 
for the dynamics of strings in the target space. These solutions usually include localization of 
perturbative string modes. For instance, extremity of open string can be restricted to move on a $p < D$ 
dimensional subspace of the target space, called a p-brane or more general solutions in the form of 
D-branes~\cite{stringbranerev}. The induced geometry on p/D-branes can be curved. In this framework, 
the observed $(3+1)$ dimensional spacetime can be a brane in $D$-dimensional fundamental target space. 
In the same manner, D-branes can be formed from condensation of closed strings, but they may be 
unstable~\cite{stringcond,stringcond0,stringcond1}. D0 branes are another class of interesting 
configuration of Yang-Mills gauge fields in the target space. They correspond to coordinate independent 
configurations, which may change with time~\cite{qgrgaugedual} or be static~\cite{qgrmatrix}. These 
models are studied in the framework of gauge-gravity duality conjecture~\cite{qgrgaugedual} and M-theory. 
These models have more common aspects with \sqgr than perturbative string models and we review them in 
more details in the next subsection.

\subsubsection{\bf M-theory and matrix theories} \label{sec:mtheorymatrix}
M-theory and matrix models are developed as candidates for non-perturbative formulation of string 
theory, see e.g.~\cite{qgrmatrixrev,qgrmatrixrev0} for review. Using various concepts, including large 
$N$ expansion of perturbative QFTs~\cite{qftlargen} and holography principle, it is conjectured that 
non-perturbative type II string theories can be described as $U(N)$ supersymmetric Yang-Mills theories 
and present quantum states of type IIA strings in D0 background~\cite{qgrgaugedual,qgrmatrix0}.

BFSS~\cite{qgrgaugedual} matrix model - called also D1+0 brane - is a 10D super Yang Mills model, 
obtained from compactification of one dimension of 11D super gravity effective field theory of string 
theory at low energy limit. It is reduced to 1+0 dimension by assuming that all the fields in the 
model are independent of 9 spatial coordinates. The dependence on the last coordinate is removed in 
what is called D0 brane or IKKT matrix model. It is demonstrated that IKKT corresponds to high 
temperature - slow variation - limit of BFSS when the Euclidean time is treated as inverse of 
temperature~\cite{qgrmatrixthermal}. 

The action of IKKT model is defined as:
\be
S_{IKKT}[X] = \frac{1}{g^2}~\text{Tr}\biggl(\frac{1}{4}[X^a,X^b][X^c,X^d] \eta_{ac} \eta_{bd} - 
\frac{i}{2}\bar{\psi}_\alpha (\mathcal{C}\sigma^a_{\alpha\beta} [A_a, \psi_\beta] \biggr ) 
\label{matrixikktaction}
\ee
where $X^a,~a = 0, \cdots, 9$ and $\psi_\alpha,~\alpha = 1, \cdots, 16$ are hermitian $N \times N$ 
matrices representing $SO(D=10)$ (Euclidean) or $SO (D-1, 1)$ (Minkowski); $\eta_{ab}$ is the 
metric of a flat Minkowski or Euclidean 10D space; $\sigma^a$'s are $16 \times 16$ Pauli matrices 
for D=10 space; and $\mathcal{C}$ is charge conjugate operator of the same dimension. The action 
$S_{IKKT}[X]$ is similar to that of type IIB superstring in Schild gauge~\cite{schilgauge}.

It is assumed that $N \rightarrow \infty$ such that $Ng^2 \equiv \bar{g}^2 < \infty$. Therefore, 
$X^a$ and $\psi_\alpha$ are respectively bosonic and fermionic $N \rightarrow \infty$ dimensional 
representations of non-Abelian $SO(D)$ (or SO(D-1,1)). Although the original motivation for this 
model has been the string theory, it con be considered without referring to strings and are also 
studied in $D \neq 10$~\cite{qgrmatrixthermal,qgrmatrixrev0}. Using variation principle, one can obtain 
{\it field} equations for $X^a$ and $\psi_\alpha$. In particular, considering only bosonic 
Yang-Mills sector, the dynamic equation for $X^a$ is:
\be
[X^a, [X^b, X^c]] \eta_{ac} = 0  \label{matrixikkteq}
\ee
Giving the finite rank of Yang-Mills symmetry group of the model and large dimension of its 
representation by $X^a$, equation (\ref{matrixikkteq}) has many solutions. Aside trivial commuting 
matrices, solutions having the form:
\be
[Y^\mu, Y^\nu] = i\theta^{\mu\nu} \mathbbm{1}, \quad \quad \mu,~\nu = 0, \cdots 2n-1, ~2n \leqslant D 
\label{moyalweyl}
\ee
where $\theta^{\mu\nu}$ is an anti-hermitian constant matrix $\theta^{\nu\mu} = -\theta^{*\mu\nu}$, 
provide reduction of space dimension and a quantized noncommutative geometry~\cite{qgrmatrixnoncommut} 
for these branes. Specifically, in case of Moyal-Weyl solution, the background 10D space is static 
with $Y^\mu = \bar{Y}^\mu, ~\mu = 0, \cdots 2n-1$ and $Y^i = 0, i = 2n, \cdots, D-1$. To define quantum 
fluctuations $Y^a$ matrices are decomposed as:
\be
Y^a = \bar{Y}^a + (A^\mu, 0) + (0, \phi^i) \label{matrxqmfluc}
\ee
Although $A^\mu$ can be considered as a $U(1)$ gauge group, it actually belongs to gravity sector and 
cannot be identified as $U(1)$ symmetry of the Standard Model. Matter fields in matrix models can arise 
like in string theory by compactifying $D-2n$ fields, see e.g.~\cite{qgrmatrixrev} for a review, or in 
case of Moyal-Weyl type solutions, by considering $k$ coinciding branes, 
see e.g.~\cite{stringbranerev,qgrmatrixym,qgrmatrixrev0} for a review. Assuming quantum superposition 
of fluctuations of $k$ branes as defined in (\ref{matrxqmfluc}), they are locally invariant under 
$SU(k)$ symmetry. Thus, in this approach $A^\mu$ and $\phi^i$ can be expanded as:
\be
A^\mu = -\theta^{\mu\nu}  A_\mu^b (\bar{Y}) T^b, \quad \quad \phi^i = \phi^i_b (\bar{Y}) T^b 
\label{matrixintersym}
\ee
where $T^b$'s are generators of adjoint representation of $SU(k)$. As indicated earlier, due to the 
fundamental non-commutation nature of $Y^a$ this construction cannot accommodate a $U(1)$ symmetry. 
In the same manner fermion fields can be constructed, but they would be in adjoint representation 
of $SU(k)$, because in 10D they are supersymmetric partners of coordinates $X^a$. For $n=2$ this 
setup lead to a quantum noncommutative $\mathbb{R}^4_\theta \subset \mathbb{R}^D$ space identified 
as the physical spacetime. 

In contrast to Randall-Sundrum type brane models, in matrix theories quantum fluctuations of geometry 
and matter do not propagate to the 10D bulk. Therefore, matrix theories, and more generally models 
based on the condensation of string modes to branes, are not dependent on the 10D background geometry. 
This solves the problem of string formulation in curved spaces discussed in Sec. \ref{sec:string}. 
But, in matrix models only fluctuations of background (target) 10D space are observables. Therefore, 
it is not clear what is the role of unobservable static (in some models) of the 10D fundamental 
background / target space. Moreover, D-branes may decay~\cite{stringcond,stringcond0,stringcond1} and 
stability of overall setup is not certain. In any case, these field/string solutions are special 
configurations, many of them are plausible~\cite{landscapecount}, and it is not clear why nature 
should prefer the one corresponding our Universe.

Finally, the low energy effective action of matrix models is a modified version of Einstein 
equation~\cite{qgrmatrixrev0, qgrmatrixinf,qgrmatrixinf0}, which is stringently constrained, specially 
with gravitational waves~\cite{qgrtest}.

\paragraph*{\bf Comparison of matrix models with \sqgr}
There are many similarities between matrix models and \sqgr, but also significant differences. In both 
models the fundamental objects are $N \rightarrow \infty$ matrices. Matrix models are pure 
(super)Yang-Mills, and $X^a$ and $\psi^i$ in (\ref{matrixikktaction}) are $N \times N$ matrices in 
adjoint representation of an {\it internal} finite rank symmetry. By contrast, \sqgr is constructed 
from a Hilbert space and includes both square and column matrices as primary entities, in adjoint and 
fundamental representations of both $\suinf$ and internal symmetries.
 
In matrix theory the large dimension of matrices is inspired by large {\it color} and loop number 
limit of QFT~\cite{qftlargen}, conjectured to present strong coupling regime. In \sqgr the motivation 
is rather cosmological and based on the observed large number of degrees of freedom in the Universe. 
These apparently different motivations converge to each other because for a perturbative estimation of 
observables, for instance $S$-matrix, up to a given degree of precision, one has to take into account 
more loops, virtual particles, and their degrees of freedom for stronger couplings. The assumption of 
\sqgr that every subsystem of the universe represents infinite degrees of freedom is an explicit 
realization of the above concept.

\sqgr uses the above axiom as a foundation for constructing other aspects of the model. Moreover, based 
on the observed spontaneous breaking and emergence of symmetries in many-body systems, see 
e.g.~\cite{manydodysymm,manydodysymm0}, it defines subsystems according to the well established criteria 
in quantum information theory~\cite{sysdiv}. The formulation of the model, specially in what concerns 
the universal quantum gravitational interaction, is completely independent of {\it internal} symmetries 
of subsystems (particles or fields), which are not constrained by the model. In fact, considering that 
when all constituents interact with each others the symmetry is $G \rightarrow \suinf$, one expects that 
many other smaller rank symmetries should have nonzero probability to arise in intermediate states, 
where the number of effectively coupled or entangled subsystems is finite.

By contrast, in matrix models the large dimension of matrices remains as a background concept and 
does not directly intervene in the construction of symmetries and dynamics. The latters are in a large 
extent inspired or concluded from superstring theory. Indeed, large matrices in these models present a 
membrane or worldsheet of a string~\cite{suninfhoppthesis,virasorosuinf,virasorosuinf0}. Despite 
the fact that matrix models can be considered as stand-alone and what they consider as fundamental 
spacetime can have other dimension than 10 of superstrings, see e.g.~\cite{qgrmatrixthermal}, BFSS 
and IKKT models and their variants are mostly constructed and studied in 10D Euclidean or Minkowski 
space. In any case, even without referring to string theory, the presence of extra-dimensions in 
matrix models is inevitable for introduction of matter and other interactions. They are considered 
to be the quantum fluctuations of a D0 condensate~\cite{qgrmatrixrev0} (and references therein) or 
compactified dimensions~\cite{qgrmatrixrev} (and references therein). However, similar to many QGR 
proposals and in contrast to \sqgr, matrix models do not provide any explanation for the observed 
dimension of spacetime.

In summary, both M-theory (matrix models) and \sqgr emphasize on the importance of $\suinf$ in a quantum 
description of gravity. But, they diverge in many details. In particular, in matrix models $\suinf$ 
symmetry is not explored. The manner internal symmetries arise in these models and constrained are 
very different. Finally, the large dimension of fundamental spacetime - the target space - in 
M-theory does not have an analogue in \sqgr.

\subsubsection{\bf Anti-de Sitter - Conformal Field Theory (AdS-CFT) duality} \label{sec:adscft}
According to AdS-CFT duality conjecture~\cite{stringgauge}, and more generally gauge-gravity 
duality~\cite{stringgauge0,stringgauge1} in M-theory, there is a one to one correspondence 
between quantum states of a suitable quantum CFT living on the boundary of a region of the spacetime and 
supergravity (string theory) in its AdS bulk. 

This conjecture is closely related to the holographic principle, but there is not yet a general proof 
for it, except in $(2+1)$D spaces~\cite{adscftentangle}. Specifically, consider a conformal field theory 
on the (1+1)D space $\mathbb{R} \times S^1$ boundary of an $AdS_3$ spacetime. Define two complementary 
subsystems $A$ and $B$ divided along $\mathbb{R}$ axis of the bulk (see figure 1 
of~\cite{adscftentangle}). The Hilbert space of the quantum CFT is factorized to 
$\hH_A \otimes \hH_B$\footnote{Notice that geometrical division to $A$ and $B$ and factorization of 
their Hilbert space is in general valid only in a given frame because as we discussed in 
Sec. \ref{sec:qgrlocality} QFT's are Type III.} and entanglement entropy between $A$ and $B$ is defined 
as $S(A) \equiv -\tr(\rho_A \ln \rho_A)$, where $\rho_A$ is the density matrix of $A$ when the state of 
$B$ is traced out.

It is proved~\cite{adscftentangle} that the static entanglement entropy, that is at $t = constant$, 
between the two subsystems is proportional to the length of the geodesic (null) curve passing inside 
the $AdS_3$ and joining the 2-point cross-section on the $t = constant$ $S^1$ boundary. More 
generally, for an $AdS_{d+2}$ spacetime the entanglement entropy is conjectured to be:
\be
S(A) = \frac{\text{Area of} \gamma_A}{4 G_N^{d+2}} \label{cftadsentropy} 
\ee
where $\gamma_A$ is the $d-$dimensional minimal (geodesic) boundary surface and $G_N$ is the Newton 
constant. Additionally, it is shown that $S(A) \rightarrow 0$ only when the size of the system goes to 
infinity~\cite{condmatterentropy,condmatterentropy0}. This case corresponds to when the two subsystems 
are infinitely separate from each other. 

We notice that the definition of subsystems in~\cite{stringgauge,adscftentangle} is geometric. This is 
an important point, because as we discussed in Sec. \ref{sec:lqftcomp}, QFTs have Type III algebra and 
Lorentz invariant quantum subsystems cannot be defined by division of their support spacetime. Thus, 
$A$ and $B$ are not properly speaking subsystems and diffeomorphism invariant. It is not clear whether 
and how this issue affects the AdS/CFT duality conjecture, specially in higher dimensional spaces for 
which a proof is not available. 

For $d=2$ the AdS $\cong R \times R \times S^d$ geometry is homomorphic to the simplest geometry of 
parameter space in \sqgr after division of the Universe to subsystems. For this case, relation with a 
CFT on the boundary in the framework of \sqgr can be understood as the following: For the whole Universe 
or an approximately isolated subsystem the size of the diffeo-surface of its $\suinf$ symmetry is 
approximately irrelevant for its observables. This property can be interpreted as an approximate 
conformal symmetry, that is scaling invariance of the parameter space of the system and its pull-back 
into the Hilbert space. Considering an external quantum clock, at a given time the parameter space of 
such an isolated subsystem is approximately 2D and its quantum dynamics is approximately a 2D CFT. Its 
operators generate a Virasoro algebra, which is a subalgebra of 
$SDiff (T^{(2)}) \cong \suinf$~\cite{virasorosuinf,virasorosuinf0,suninftorus}. Invariance by scaling 
means that any arbitrary diffeomorphism can be bring back to a surface preserving one.

\section{\bf Outline}  \label{sec:outline}
Comparison of several popular QGR models with \sqgr proposal in this work finds a number of common or 
analogous features between them. We discussed the origin of these properties and showed that they arise 
in \sqgr either from its axioms or can be concluded from them. Giving simple axioms and systematic and 
natural emergence of common features in \sqgr, this model may help to clarify some of puzzling 
properties of other QGR models.

\sqgr is not yet applied to physical phenomena in which QGR may be involved, such as:
\begin{description}
\item Hawking radiation;
\item Spacetime singularities;
\item Information loss paradox of black holes;
\item Topology of parameter space of the model identified as the classical spacetime and whether it 
can be changed;
\item Particle physics at Planck scale.
\end{description}
Regarding the last item in this list, as mentioned in Sec. \ref{secevolsub}, we expect more unbroken 
gauge symmetries at high energies, which can be also interpreted as the internal symmetry $G$. 
Therefore, a crucial task is to determine how internal symmetries vary with energy scale $\Lambda$. 
Particle physics experiments show that $G |_{\Lambda \sim 1 \text{TeV}} = SU(3) \times SU(2) \times U(1)$, 
i.e. the Standard Model symmetry. But, signature of interactions at higher energies may be smeared by 
the physics at lower energies~\cite{houribranecr}. Nonetheless, if many-body high energy states behave 
similar to low energy many-body systems, the analogy may help to find best criteria for detecting 
signatures of {\it phase transitions} due to symmetry transition at high energies in (astro-)particle 
physics experiments or cosmological observations. Other topics in the above list would be the subject 
of future investigations.

\appendix
\section {Classical limit of $\suinf$ Yang-Mills of subsystems} \label{app}
We first review some of properties of curvatures of (pseudo)Riemannian manifolds. For any Riemannian 
or pseudo-Riemannian manifold $\mm,g)$ of dimension $d \geqslant 2$ equipped with a Levi-Civita 
connection $\nabla$ the Riemann curvature (1,3) tensor at point $p \in \mm$ is defined as:
\be
R_p (X, Y)Z = \nabla_X \nabla_Y - \nabla_Y \nabla_X - \nabla_{[X,Y]}  \label{riemanncurve13}
\ee
Vector fields $X,~Y,~Z \in T\mm_p$ and $T\mm_p$ is the tangent space at $p$. When $X,~Y,~Z$ are chosen 
to be $\partial_i \equiv \partial / \partial x^i$ basis of the tangent space for coordinates 
$x^i,~ i=0, \cdots d-1$, one recover the usual coordinate dependent definition of the Riemann curvature 
tensor (we drop $p$ because it corresponds to the point with coordinates $x^i$:
\bea
&& R (\partial_i, \partial_j) \partial_k = R^{l}_{kij}, \label{riemanncurve13c} \\
&& R_{ijkl} \equiv R(\partial_i, \partial_j, \partial_k, \partial_l) \equiv \bigg \langle R (\partial_i, 
\partial_j) \partial_k, \partial_l \bigg \rangle = g_{ml} R^m_{kij} \label{riemanncurve04c}
\eea
Using the notation defined in (\ref{riemanncurve04c}) for (0,4) Riemann curvature tensor sectional 
curvature $K(\Pi) = K(X, Y)$ with respect to a 2D plane $\Pi \subset T\mm_p$ containing two vectors  
$X, Y \in T\mm_p$ at $p \in \mm$ is defined as:
\be
K(\Pi) \equiv K(X,Y) = \frac{R_p(X, Y, X, Y)}{\langle X, X\rangle \langle Y, Y\rangle - 
\langle X, Y\rangle^2}  \label{sectionalcurve}
\ee
Notice that $K(\Pi)$ is independent of the choice of $X$ and $Y$ and depends only of the plane passing 
through them. It can be shown that $\langle R_p (X, Y)Z, W)\rangle$ can be expanded with respect to 
sectional curvatures~\cite{seccurvproof}. Using relations between different curvature tensors of a 
Riemannian manifold, Ricci scalar at point $p \in \mm$ is defined as~\cite{seccurvproof0}:
\be
R(p) = \sum_{i \neq j} R_p (e_i, e_j, e_i, e_j) = \sum_{i \neq j} K_p (e_i, e_j) \label{riccicurvsec}
\ee
where $e_i, ~ i=0, \cdots, d-1$ is an orthonormal basis of $T\mm_p$. From (\ref{riccicurvsec}) we 
conclude that there is only one sectional curvature at each point of a 2D surface and it is equal to 
its Ricci scalar curvature.

In order to extend the relation (\ref{eulerint}) between $\suinf$ Yang-Mills action and Ricci scalar 
curvature of its diffeo-surface of an isolated quantum system with $\suinf$ symmetry to large number of 
such systems when the Universe is divided to subsystems, we have to integrate over their contribution. 
In Sec. \ref {secsubparam} we showed that the parameter space of subsystems is (3+1)D dimensional. 
Applying definitions of curvature tensors to this parameter space, each sectional curvature in 
(\ref{riccicurvsec}) can be interpreted as Ricci curvature of the diffeo-surface of a subsystem and 
the summation in the r.h.s. of (\ref{riccicurvsec}) and integration over the volume of the parameter 
space amounts to taking into account all subsystems of the Universe. Therefore, the r.h.s. of 
(\ref{r4}) corresponds to the r.h.s. of (\ref{eulerint}) when the Universe is divided to subsystems. 
In the same way, the l.h.s. of (\ref{r4}) corresponds to the l.h.s. of (\ref{eulerint}) when the 
contribution of subsystems are calculated separately. However, (\ref{r4}) is valid only in the 
classical limit, because as discussed in Sec. \ref{sec:geom}, after the selection of reference and 
clock the ensemble of remaining subsystems should be considered as an open quantum system. Therefore, 
(\ref{r4}) is an approximation, valid only in the classical limit where the reference and the clock can 
be considered as classical.


\begin{thebibliography}{999}
\bibitem {houriqmsymmgr} \Name{H.}{Ziaeepour} ~ Making a Quantum Universe: Symmetry and Gravity ~ \Journal{\MDU}{}{2021}{-}, \href{https://www.mdpi.com/2218-1997/6/11/194}{Special Issue 80 Years of Professor Wigner's Seminal Work "On Unitary Representations of the Inhomogeneous Lorentz Group"}, \href{}{[arXiv:2009.03428]}.
\bibitem {grinconsist} \Name{K.}{Eppley} ; \Name{E.}{Hanna} ~ The Necessity of Quantizing the Gravitational Field. \Journal{\FOP}{7}{1977}{51}.
\bibitem {greos} \Name{T.}{Jacobson} ~ Thermodynamics of Spacetime: The Einstein Equation of State. \Journal{\PRL}{75}{1995}{1260}, \href{http://arxiv.org/abs/gr-qc/9504004}{[arXiv:gr-qc/9504004]}. 
\bibitem {hologprin} \Name{J.D.}{Bekenstein} ~ Statistical black-hole thermodynamics. \Journal{\PRD}{12}{1975}{3077}.
\bibitem {hologprin0} \Name{G.}{t'Hooft} ~ Dimensional Reduction in Quantum Gravity. (1993), \href{https://arxiv.org/abs/gr-qc/9310026}{[arXiv:gr-qc/9310026]}.
\bibitem {hologprin1} \Name{L.}{Susskin} ~ The World as a Hologram. \Journal{JMP}{36}{1995}{6377}, \href{https://arxiv.org/abs/hep-th/9409089}{[arXiv:hep-th/9409089]}.
\bibitem {hologrev} \Name{R.}{Bousso} ~ The holographic principle. \Journal{\RMP}{74}{2002}{825}, \href{https://arxiv.org/abs/hep-th/0203101}{[arXiv:hep-th/020310]}.
\bibitem {condmatterentropy} \Name{C.}{Holzhey} ; \Name{F.}{Larsen} ; \Name{F.}{Wilczek} ~ Geometric and Renormalized Entropy in Conformal Field Theory. \Journal {\NPB}{424}{1994}{443}, \href{https://arxiv.org/abs/hep-th/9403108}{[arXiv:hep-th/9403108]}. 
\bibitem {condmatterentropy0} \Name{P.}{Calabrese} ; \Name{J.}{Cardy} ~ Entanglement Entropy and Quantum Field Theory. \Journal {\JSM}{0406}{2004}{P06002}, \href{https://arxiv.org/abs/hep-th/0405152}{[arXiv:hep-th/0405152]}.
\bibitem {hourivacuum} \Name{H.}{Ziaeepour} ~ Issues with vacuum energy as the origin of dark energy. \Journal{\MPL}{27}{2012}{1250154}, \href{http://arxiv.org/abs/arXiv:1205.3304}{]arXiv:1205.3304]}.
\bibitem {qmwithouttime} \Name{C.}{Rovelli} ~ Quantum mechanics without time: A model. \Journal{\PRD}{42}{1990}{2638}.
\bibitem {qgrspaceless} \Name{F.}{Markopoulou} ~ Space does not exist, so time can. (2009), \href{https://arxiv.org/abs/0909.1861}{[arXiv:0909.1861]}.
\bibitem {qgrentangle} \Name{M.}{Van Raamsdonk} ~ Building up spacetime with quantum entanglement. \Journal {\GRG}{42}{2010}{2323}, \Journal{\GRG}{42}{2010}{2323}; \Journal {\IMD}{19}{2010}{2429}, \href{http://arxiv.org/abs/1005.3035}{[arXiv:1005.3035]}.
\bibitem {qgrentangle1} \Name{C.}{Cao} ; \Name{S.M.}{Carroll} ; \Name{S.}{Michalakis} ~ Space from Hilbert Space: Recovering Geometry from Bulk Entanglement. \Journal{\PRD}{95}{2017}{024031}, \href{http://arxiv.org/abs/1606.08444}{[arXiv:1606.08444]}.
\bibitem {qmtimepage} \Name{D.N.}{Page} ; \Name{W.K.}{Wootters} ~ Evolution without evolution: Dynamics described by stationary observables. \Journal{\PRD}{27}{1983}{2885}.
\bibitem {houriqmsymmeos} In preparation (2021).
\bibitem {qgrlocalqm1} \Name{S.B.}{Giddings} ~ Quantum-first gravity. \Journal {\FOP}{49}{2019}{177}, \href{http://arxiv.org/abs/1803.04973}{[arXiv:1803.04973]}.
\bibitem {qgrlocalqm} \Name{S.B.}{Giddings} ~ Hilbert space structure in quantum gravity: an algebraic perspective. \Journal {\JHE}{2015}{2015}{1}, \href{https://arxiv.org/abs/1503.08207}{[arXiv:1503.08207]}
\bibitem {bhentropy} \Name{J.D.}{Bekenstein} ~ Black Holes and Entropy. \Journal {\PRD}{7}{1973}{2333}.
\bibitem {hawkingrad} \Name{S.}{Hawking} ~ Break of predictability in gravitational collapse. \Journal{\PRD}{14}{1976}{246}.
\bibitem {bhentropyloss} \Name{W.H.}{Zurek} ~ Entropy Evaporated by a Black Hole. \Journal{\PRL}{49}{1982}{1683}.
\bibitem {qgrentangle2} \Name{C.}{Cao} ; \Name{S.M.}{Carroll} ~ Bulk Entanglement Gravity without a Boundary: Towards Finding Einstein's Equation in Hilbert Space. \Journal{\PRD}{97}{2018}{086003}, \href{https://arxiv.org/abs/1712.02803}{[arXiv:1712.02803]}.
\bibitem {qgrcanonical} \Name{B.}{Dewitt} ~ Quantum Theory of Gravity. I. The Canonical Theory. \Journal {\PRV}{160}{1967}{1113}.
\bibitem {qgrcanonical0} \Name{J.B.}{Hartle} ; \Name{S.W.}{Hawking} ~ Wave function of the Universe.  \Journal{\PRD}{28}{1983}{2960}. 
\bibitem {qgrgeometrody} \Name{J.A.}{Wheeler} ~ On the nature of quantum geometrodynamics. \Journal {\APH} {2}{1957}{604}.
\bibitem {qgrearlyhist} \Name{A.}{Rocci} ~ On first attempts to reconcile quantum principles with gravity. \Journal{\JPC}{470}{2013}{012004}, \href{http://arxiv.org/abs/1309.7336}{[arXiv:1309.7336]}.. 
\bibitem {qgrgeometrodrev} \Name{C.}{Kiefer} ~ Quantum geometrodynamics: Whence, whither? \Journal {\GRG}{41}{2009}{877}, \href{http://arxiv.org/abs/0812.0295}{[arXiv:0812.0295]}.
\bibitem {admgr} \Name{R.}{Arnowitt} ; \Name{S.}{Deser} ; \Name{C.}{Misner} ~ Dynamical Structure and Definition of Energy in General Relativity. \Journal{\PRV}{116}{1959}{1322}, \href{https://arxiv.org/abs/gr-qc/0405109}{[arXiv:gr-qc/0405109]} 
\bibitem {sysdiv} \Name{P.}{Zanardi} ; \Name{D.}{Lidar} ; \Name{S.}{Lloyd} ~ Quantum tensor product structures are observable-induced. \Journal{\PRL}{92}{2004}{060402}, \href{http://arxiv.org/abs/quant-ph/0308043}{[arXiv:quant-ph/0308043]}.
\bibitem {stringgauge} \Name{J.M.}{Maldacena} ~ The Large N Limit of Superconformal Field Theories and Supergravity. \Journal {\ATM}{2}{1998}{231}, \href{http://arxiv.org/abs/hep-th/9711200}{[arXiv:hep-th/9711200]}.
\bibitem {stringgauge0} \Name{E.}{Witten} ~ Anti De Sitter Space And Holography. \Journal{\ATM}{2}{1998}{253}, \href{https://arxiv.org/abs/hep-th/9802150}{[arXiv:hep-th/9802150]}.
\bibitem {stringgauge1} \Name{O.}{Aharony} ; \Name{S.S.}{Gubser} ; \Name{J.}{Maldacena} ; \Name{H.}{Ooguri} ; \Name{Y.}{Oz} ~ Large N Field Theories, String Theory and Gravity. \Journal{\PRE}{323}{2000}{183}, \href{https://arxiv.org/abs/hep-th/9905111}{[arXiv:hep-th/9905111]}.
\bibitem {qmdirac} Dirac, P.A.M.: The Principles of Quantum Mechanics, Oxford University Press (1958).
\bibitem {qmvonneumann} Von Neumann, J.: Mathematical Foundation of Quantum Theory, Princeton University Press, (1955).
\bibitem {suninfhoppthesis} \Name{J.}{Hoppe} ~  Quantum Theory of a Massless Relativistic Surface and a Two-dimensional Bound State Problem. Ph.D. Thesis, MIT, Cambridge, MA, USA, (1982).
\bibitem {suninfym} \Name{E.G.}{Floratos} ; \Name{J.}{Iliopoulos} ; \Name{G.}{Tiktopoulos} ~ A note on $SU(\infty)$ classical Yang-Mills theories. \Journal{\PLB}{217}{1989}{285}.
\bibitem {suninftorus} \Name{J.}{Hoppe} ~ Diffeomorphism Groups, Quantization, and $SU(\infty)$ . \Journal {\IMA}{4}{1989}{5235}. 
\bibitem {suninfrep} \Name{J.}{Hoppe} ; \Name{P.}{Schaller} ~ Infinitely Many Versions of $SU(\infty)$. \Journal {\PLB}{237}{1990}{407}.
\bibitem {suninfrep0} \Name{Y.}{Zunger} ~ Why Matrix theory works for oddly shaped membranes. \Journal{\PRD}{64 }{2001}{086003}, \href{http://arxiv.org/abs/hep-th/0106030}{[arXiv:hep-th/0106030]}.
\bibitem {houriqgr} \Name{H.}{Ziaeepour} ~ And what if gravity is intrinsically quantic ? \Journal{\JPC}{174}{2009}{012027} ~ \href{http://arxiv.org/abs/0901.4634}{[arXiv:0901.4634]}.
\bibitem {cartandecomp} \Name{Z-Y.}{Su} ~ A Scheme of Cartan Decomposition for su(N). (2006), \href{http://arxiv.org/abs/quant-ph/0603190}{[arXiv:quant-ph/0603190]}.
\bibitem {houriqmsymm} \Name{H.}{Ziaeepour} ~ Foundational role of symmetry in Quantum Mechanics and Quantum Gravity. in ``Quantum Mechanics: Theory, Analysis, and Applications'', Nova Science Publishers Inc., New York (2019), \href{http://arxiv.org/abs/arXiv:1305.4349}{[arXiv:1305.4349]}.
\bibitem {houriqmsymm0} \Name{H.}{Ziaeepour} ~ Symmetry as a foundational concept in Quantum Mechanics. \Journal{\JPC}{626}{2015}{012074}, \href{http://arxiv.org/abs/arXiv:1502.05339}{[arXiv:1502.05339]}.
\bibitem {qmtimedef} \Name{P.A.}{Hoehn} ; \Name{A.R.H.}{Smith} ; \Name{M.P.E.}{Lock} ~ The Trinity of Relational Quantum Dynamics. (2019), \href{http://arxiv.org/abs/1912.00033}{[arXiv:1912.00033]}.
\bibitem {qmspeed} \Name{L.}{Mandelstam} ; \Name{I.}{Tamm} ~ The Uncertainty Relation Between Energy and Time in Non-relativistic Quantum Mechanics. \Journal{\JPU}{9}{1945}{249}.
\bibitem {qmrefsubsys} \Name{Ph}{Hoehen} ; \Name{M.P.E.}{Lock} ; \Name{S.}{Ali Ahmad} ; \Name{A.R.H.}{Smith} ; \Name{T.D.}{Galley} ~ Quantum Relativity of Subsystems. (2021), \href{https://arxiv.org/abs/2103.01232}{[arXiv:2103.01232]}.
\bibitem {qgrlagrangian} \Name{L.}{Rosenfeld} ~ Zur Quantelung der Wellenfelder. \Journal{\em Annal der Physik}{397}{1930}{113}.
\bibitem {qgrcanonical} \Name{B.}{Dewitt} ~ Quantum Theory of Gravity. I. The Canonical Theory. \Journal {\PRV}{160}{1967}{1113}.
\bibitem {qgrcanonical0} \Name{J.B.}{Hartle} ; \Name{S.W.}{Hawking} ~ Wave function of the Universe. \Journal{\PRD}{28}{1983}{2960}. 
\bibitem {qgrearlyhist} \Name{A.}{Rocci} ~ On first attempts to reconcile quantum principles with gravity. \Journal{\JPC}{470}{2013}{012004}, \href{http://arxiv.org/abs/1309.7336}{[arXiv:1309.7336]}.. 
\bibitem {curvatureregge} \Name{T.}{Regge} ~ General Relativity without Coordinates. \Journal{\NCE}{19}{1961}{558}.
\bibitem {reggecal} \Name{R.}{Gambini} ; \Name{J.}{Pullin} ~ Consistent discretization and canonical classical and quantum Regge calculus. \Journal {\IMD}{15}{2006}{1699}, \href{https://arxiv.org/abs/gr-qc/0511096}{[arXiv:gr-qc/0511096]}.
\bibitem {qgrponzanoreggge} \Name{G.}{Ponzano}, \Name{T.}{Regge} ~ Semiclassical limit of Racah coefficients. p1-58; in Spectroscopic and group theoretical methods in physics, ed. F. Bloch, North-Holland Publ. Co. Amsterdam, (1968).
\bibitem {ashtekarvar} \Name{A.}{Ashtekar} ~ New Variables for Classical and Quantum Gravity. \Journal{\PRL}{57}{1986}{2244}.
\bibitem {lqgrreggerev} \Name{G.}{Immirzi} ~ Quantum Gravity and Regge Calculus. \Journal{\NPBS}{57}{1997}{65}, \href{https://arxiv.org/abs/gr-qc/9701052}{[arXiv:gr-qc/9701052]}.
\bibitem {lqgrev} \Name{C.}{Rovelli} ~ \emph{Quantum Gravity}; Cambridge University Press:  Cambridge, UK, (2004).
\bibitem {lqgrev0} \Name{A.}{Ashtekar} ; \Name{J.}{Lewandowski} ~ Background Independent Quantum Gravity: A Status Report.  \Journal{\CQG}{21}{2004}{R53}, \href{http://arxiv.org/abs/gr-qc/0404018}{[arXiv:gr-qc/0404018]}.
\bibitem {immirziparam} \Name{G.}{Immirzi} ~ Real and complex connections for canonical gravity. \Journal{\CQG}{14}{1997}{L177}, \href{https://arxiv.org/abs/gr-qc/9612030}{[arXiv:gr-qc/9612030]}.
\bibitem {qgrspinfoam} \Name{J.W.}{Barrett} ; \Name{L.}{Crane} ~ Relativistic spin networks and quantum gravity. \Journal{\JMP}{39}{1998}{3296}, \href{https://arxiv.org/abs/gr-qc/9709028}{[arXiv:gr-qc/9709028]}.
\bibitem {qgrspinfoam0} \Name{J.W.}{Barrett} ; \Name{L.}{Crane} ~ A Lorentzian Signature Model for Quantum General Relativity. \Journal {\CQG}{17}{2000}{3101}, {[arXiv:gr-qc/9904025]}.
\bibitem {lqgfoam} \Name{E.R.}{Livine} ~ Projected Spin Networks for Lorentz connection: Linking Spin Foams and Loop Gravity. \Journal{\CQG}{19}{2002}{5525}, \href{http://arxiv.org/abs/gr-qc/0207084}{[arXiv:gr-qc/0207084]}.
\bibitem {lqgareaop} \Name{A.}{Ashtekar} ; \Name{C.}{Rovelli} ; \Name{L.}{Smolin} ~ Weaving a Classical Metric with Quantum Threads. \Journal {\PRL}{69}{1992}{237}.
\bibitem {lqgtetrahed} \Name{C.}{Rovelli} ; \Name{D.}{Colosi} ; \Name{L.}{Doplicher} ; \Name{W.}{Fairbairn}, \Name{L.}{Modesto}, \Name{Karim}{Noui} ~ Background independence in a nutshell. \Journal{\CQG}{22}{2005}{2971}, \href{https://arxiv.org/abs/gr-qc/0408079}{[arXiv:gr-qc/0408079]}.
\bibitem {lqghilbert} \Name{L.}{Smolin} ~ An invitation to loop quantum gravity. in ~ Quantum Theory and Symmetries. Ed. {P.C.}{Argyres}, \etal, World Scientific (2004), \href{https://arxiv.org/abs/hep-th/0408048}{[arXiv:hep-th/0408048]}.
\bibitem {lqgfoamreality} \Name{S.K.}{Maran} ~ Reality Conditions for Spin Foams. (2005), \href{https://arxiv.org/abs/gr-qc/0511014}{[arXiv:gr-qc/0511014]}.
\bibitem {lqglorentzviol} \Name{J.}{Collins} ; \Name{A.}{Perez} ; \Name{D.}{Sudarsky} ; \Name{L.}{Urrutia} ; \Name{H.}{Vucetich} ~ Lorentz invariance and quantum gravity: an additional fine-tuning problem? \Journal {\PRL}{93}{2004}{191301}, \href{https://arxiv.org/abs/gr-qc/0403053}{[arXiv:gr-qc/0403053]}.
\bibitem {lggcovar} \Name{R.}{Gambini} ; \Name{J.}{Pullin} ~ Emergent diffeomorphism invariance in a discrete loop quantum gravity model. \Journal {\CQG}{26}{2009}{035002}, \href{https://arxiv.org/abs/0807.2808}{[arXiv:0807.2808]}.
\bibitem {lggcovar0} \Name{A.}{Ashtekar} ~ Some surprising implications of background independence in canonical quantum gravity. \Journal {\GRG}{41}{2009}{1927}, \href{https://arxiv.org/abs/0904.0184}{[arXiv:0904.0184]}.
\bibitem {lqggwdispersion} \Name{M.}{Bojowald} ; \Name{G.}{Mortuza Hossain} ~ Loop quantum gravity corrections to gravitational wave dispersion. \Journal {\PRD}{77}{2008}{023508}, \href{https://arxiv.org/abs/0709.2365}{[arXiv:0709.2365]}.
\bibitem {lqgphenomnrev} \Name{F.}{Girelli}, \Name{F.}{Hinterleitner}, \Name{S.A.}{Major} ~ Loop Quantum Gravity Phenomenology: Linking Loops to Observational Physics. \Journal {\em SIGMA}{8}{2012}{098}, \href{https://arxiv.org/abs/1210.1485}{[arXiv:1210.1485]}.
\bibitem {grestgrb090510a} \Name{A.A.}{Abdo} ; \Name{M.}{Ackermann} ; \Name{M.}{Ajello} ;\Name{K.}{Asano} ; \Name{W.B.}{Atwood} ; \Name{M.}{Axelsson} ; \Name{L.}{Baldini} ; \Name{J.}{Ballet} ; \Name{G.}{Barbiellini} ; \Name{M.G.}{Baring} ; \etal ~ A limit on the variation of the speed of light arising from quantum gravity effects. \Journal {\NAT}{462}{2009}{331}, \href{http://arxiv.org/abs/0908.1832}{[arXiv:0908.1832]}.
\bibitem {gwdisperligo} The LIGO Scientific Collaboration ~ Tests of General Relativity with GW170817. \Journal{\PRL}{123}{2019}{011102}, \href{https://arxiv.org/abs/1811.00364}{[arXiv:1811.00364]}.
\bibitem {gwdisperligo0} The LIGO Scientific Collaboration ~ Tests of General Relativity with the Binary Black Hole Signals from the LIGO-Virgo Catalog GWTC-1. \Journal {\PRD}{100}{2019}{104036}, \href{https://arxiv.org/abs/1903.04467}{[arXiv:1903.04467]}.
\bibitem {lqgimmirzfermion} \Name{A.}{Perez}, \Name{C.}{Rovelli} ~ Physical effects of the Immirzi parameter. \Journal {\PRD}{73}{2006}{044013}, \href{https://arxiv.org/abs/gr-qc/0505081}{[arXiv:gr-qc/0505081]}.
\bibitem {grfifthconstraint} \Name{J.}{Berg\'e} ; \Name{M.}{Pernot-Borr\`as} ; \Name{J.-P.}{Uzan} ; \Name{P.}{Brax} ; \Name{R.}{Chhun} ; \Name{G.}{Métris} ; \Name{M.}{Rodrigues} ; \Name{P.}{Touboul} ~ MICROSCOPE's constraint on a short-range fifth force. (2021), \href{https://arxiv.org/abs/2102.00022}{[arXiv:2102.00022]}.
\bibitem {lqghistories} \Name{M.}{Gaul} ; \Name{C.}{Rovelli} ~ Loop Quantum Gravity and the Meaning of Diffeomorphism Invariance. \Journal {\LNP}{541}{2000}{277}, \href{https://arxiv.org/abs/gr-qc/9910079}{[arXiv:gr-qc/9910079]}.
\bibitem {lqghistories0} \Name{J.J.}{Halliwell} ; \Name{P.}{Wallden} ~ Invariant Class Operators in the Decoherent Histories Analysis of Timeless Quantum Theories. \Journal{\PRD}{73}{2006}{024011}, \href{https://arxiv.org/abs/gr-qc/0509013}{[arXiv:/0509013]}.
\bibitem {lqgtime} \Name{M.}{Reisenberger} ; \Name{C.}{Rovelli} ~ Spacetime states and covariant quantum theory. \Journal {\PRD}{65}{2002}{125016}, \href{https://arxiv.org/abs/gr-qc/}{[arXiv:gr-qc/0111016]}.
\bibitem {lqgcoarse} \Name{D.R.}{Terno} ~ Quantum information in loop quantum gravity. \Journal{\JPC}{33}{2006}{469}, \href{https://arxiv.org/abs/gr-qc/0512072}{[arXiv:gr-qc/0512072]}.
\bibitem {lqgphase} \Name{K.}{Giesel} ; \Name{T.}{Thiemann} ~ Algebraic Quantum Gravity (AQG) IV. Reduced Phase Space Quantisation of Loop Quantum Gravity. \Journal{\CQG}{27}{2010}{175009}, \href{https://arxiv.org/abs/0711.0119}{[arXiv:0711.0119]}.
\bibitem {lqgphase0} \Name{V.}{Husain} ; \Name{T.}{Pawlowski} ~ Time and a physical Hamiltonian for quantum gravity. \Journal{\PRL}{108}{2012}{141301}, \href{https://arxiv.org/abs/1108.1145}{[arXiv:1108.1145]}. 
\bibitem {lqgphase1} \Name{K.}{Giesel} ; \Name{A.}{Vetter} ~ Reduced Loop Quantization with four Klein-Gordon Scalar Fields as Reference Matter. (2016), \href{https://arxiv.org/abs/1610.07422}{[arXiv:1610.07422]}.
\bibitem {lqgmatter} \Name{S.}{Gielen} ; \Name{D.}{Oriti} ~ Cosmological perturbations from full quantum gravity. \Journal{\PRD}{98}{2018}{106019}, \href{https://arxiv.org/abs/1709.01095}{[arXiv:1709.01095]}.
\bibitem {spaceemerge} \Name{F.}{Wilczek} ~ Riemann-Einstein Structure from Volume and Gauge Symmetry. \Journal {\PRL}{80}{1998}{4851}, \href{https://arxiv.org/abs/hep-th/9801184}{[arXiv:hep-th/9801184]}.
\bibitem {spaceemerge0} \Name{N.}{Seiberg} ~ Emergent Spacetime. in ~ The Quantum Structure of Space and Time. Ed: J. Harvey, p. 163 World Scientific (2007), \href{https://arxiv.org/abs/hep-th/0601234}{[arXiv:hep-th/0601234]}.
\bibitem {spaceemerge1} \Name{H.}{Westman} ; \Name{S.}{Sonego} ~ Coordinates, observables and symmetry in relativity. \Journal{\APH}{324}{2009}{1585} \href{https://arxiv.org/abs/0711.2651}{[arXiv:0711.2651]}.
\bibitem {qgrgaugesep} \Name{F.}{Wilczek} ~ Riemann-Einstein Structure from Volume and Gauge Symmetry. \Journal {\PRL}{80}{1998}{4851}, \href{http://arxiv.org/abs/hep-th/9801184}{[arXiv:hep-th/9801184]}.
\bibitem {qgrgaugesep0} \Name{A.}{Torres-Gomez} ; \Name{K.}{Krasnov} ~ Gravity-Yang-Mills-Higgs unification by enlarging the gauge group. \Journal{\PRD}{81}{2010}{085003}, \href{http://arxiv.org/abs/0911.3793}{[arXiv:0911.3793]}. 
\bibitem {qgrgaugesep1} \Name{J.W.}{Barrett} ; \Name{S.}{Kerr} ~ Gauge gravity and discrete quantum models. (2013), \href{http://arxiv.org/abs/1309.1660}{[arXiv:1309.1660]}. 
\bibitem {qgrthermo} \Name{T.}{Padmanabhan} ~ Gravity and the Thermodynamics of Horizons. \Journal{\PRE}{406}{2005}{49}, \href{https://arxiv.org/abs/gr-qc/0311036}{[arXiv:gr-qc/0311036]}.
\bibitem {qgrthermo0} \Name{T.}{Padmanabhan} ~ Gravity as an emergent phenomenon: A conceptual description. 
\Journal {\AIP}{939}{2007}{114}, \href{https://arxiv.org/abs/0706.1654}{[arXiv:0706.1654]}.
\bibitem {qgrthermo1} \Name{E.P.}{Verlinde} ~ On the Origin of Gravity and the Laws of Newton. \Journal{\JHE}{1104}{2011}{029}, \href{https://arxiv.org/abs/1001.0785}{[1001.0785]}.
\bibitem {qgrhistory} \Name{J.B.}{Hartle} ~ Generalizing Quantum Mechanics for Quantum Spacetime. in ~ The Quantum Structure of Space and Time: ed. by D. Gross, M. Henneaux, A. Sevrin, World Scientific, Singapore, (2007), \href{https://arxiv.org/abs/gr-qc/0602013}{[arXiv:gr-qc/0602013]}.
\bibitem {qgrhistory0} \Name{S.B.}{Giddings} ~ Universal quantum mechanics. \Journal{\PRD}{78}{2008}{084004}, \href{https://arxiv.org/abs/0711.0757}{[arXiv:0711.0757]}.
\bibitem {qmhistories} \Name{R.B.}{Griffiths} ~ Consistent Histories and the Interpretation of Quantum Mechanics. \Journal{\JSP}{36}{1984}{219}.
\bibitem {qmhistories0} \Name{C.J.}{Isham} ~ Quantum Logic and the Histories Approach to Quantum Theory. \Journal{\JMP}{35}{1994}{2157}, \href{http://arxiv.org/abs/gr-qc/9308006}{[arXiv:gr-qc/9308006]}.
\bibitem {qmcurve} \Name{N.D.}{Birrell} ; \Name{P.C.W.}{Davies} ~ Quantum Fields in Curved Space. Cambridge University Press: Cambridge, UK, (1982).
\bibitem {qgrhistoryrev} \Name{J.}{Henson} ~ Quantum Histories and Quantum Gravity. \Journal{J.Phys.Conf.Ser}{174}{2009}{012020}, \href{https://arxiv.org/abs/0901.4009}{[arXiv:0901.4009]}.
\bibitem {qgrhistorymultivers} \Name{J.B.}{Hartle} ~ Quantum Multiverses. \emph{arXiv} \textbf{(2018)}, \href{https://arxiv.org/abs/1801.08631}{[arXiv:1801.08631]}.
\bibitem {qgrlocalqm0} \Name{W.}{Donnelly} ; \Name{S.B.}{Giddings} ~ How is quantum information localized in gravity? \Journal{\PRD}{96}{2017}{086013}, \href{https://arxiv.org/abs/1706.03104}{[arXiv:1706.03104]}.
\bibitem {qgrlocalqm2} \Name{W.}{Donnelly} ; \Name{S.B.}{Giddings} ~ Gravitational splitting at first order: Quantum information localization in gravity. \Journal{\PRD}{98}{2018}{086006}, \href{https://arxiv.org/abs/1805.11095}{[arXiv:1805.11095]}.
\bibitem {vonNeumannoptype} \Name{J.}{von Neumann} ~ Mathematische Grundlagen der Quantunmechanik, Springer, Berlin (1932).
\bibitem {type3op} \Name{J.}{Yngvason} ~ The Role of Type III Factors in Quantum Field Theory. \Journal{\RMA}{55}{2005}{135}, \href{https://arxiv.org/abs/math-ph/0411058}{[math-ph/0411058]}.
\bibitem {qgrgaugedual} \Name{T.}{Banks} ; \Name{W.}{Fischler} ; \Name{S.H.}{Shenker} ; \Name{L.}{Susskind}  ~ M Theory As A Matrix Model: A Conjecture. \Journal{\PRD}{55}{1997}{5112}, \href{https://arxiv.org/abs/hep-th/9610043}{[arXiv:hep-th/9610043]}.
\bibitem {qgrmatrix} \Name{N.}{Ishibashi} ; \Name{H.}{Kawai} ; \Name{Y.}{Kitazawa} ; \Name{A.}{Tsuchiya} ~ A Large-N Reduced Model as Superstring. \Journal{\NPB}{498}{1997}{467}, \href{http://arxiv.org/abs/hep-th/9612115}{[arXiv:hep-th/9612115]}.
\bibitem {qgrgaugedual0} \Name{O.}{Aharony} ; \Name{S.S.}{Gubser} ; \Name{J.}{Maldacena} ; \Name{H.}{Ooguri} ; \Name{Y.}{Oz} ~ Large N Field Theories, String Theory and Gravity. \Journal{\PRE}{323}{2000}{183}, \href{https://arxiv.org/abs/hep-th/9905111}{[hep-th/9905111]}.
\bibitem {qgrgaugesep} \Name{F.}{Wilczek} ~ Riemann-Einstein Structure from Volume and Gauge Symmetry. \Journal {\PRL}{80}{1998}{4851}, \href{http://arxiv.org/abs/hep-th/9801184}{[arXiv:hep-th/9801184]}.
\bibitem {qgrgaugesep0} \Name{A.}{Torres-Gomez} ; \Name{K.}{Krasnov} ~ Gravity-Yang-Mills-Higgs unification by enlarging the gauge group. \Journal{\PRD}{81}{2010}{085003}, \href{http://arxiv.org/abs/0911.3793}{[arXiv:0911.3793]}. 
\bibitem {qgrgaugesep1} \Name{J.W.}{Barrett} ; \Name{S.}{Kerr} ~ Gauge gravity and discrete quantum models. \emph{arXiv} \textbf{(2013)}, \href{http://arxiv.org/abs/1309.1660}{[arXiv:1309.1660]}. 
\bibitem {qgrautomatom} \Name{G.}{'t Hooft} ~ Dimensional Reduction in Quantum Gravity. (1993), \href{http://arxiv.org/abs/gr-qc/9310026}{[arXiv:gr-qc/9310026]}.
\bibitem {adscftentangle} \Name{S.}{Ryu} ; \Name{T.}{Takayanagi} ~ Holographic Derivation of Entanglement Entropy from AdS/CFT. \Journal{\PRL}{96}{2006}{181602}, \href{https://arxiv.org/abs/hep-th/0603001}{[arXiv:hep-th/0603001]}.
\bibitem {adscftrev} \Name{J.}{Maldacena} ~ The gauge/gravity duality. in ~ Black Holes in Higher Dimensions Ed. G. Horowitz, Cambridge University Press, (2012) \href{https://arxiv.org/abs/1106.6073}{[arXiv:1106.6073]}.
\bibitem {stringrev} \Name {M.B.}{Green} ; \Name {J.H.}{Schwarz} ; \Name {E.}{Witten} ~ \emph{Superstring Theory I \& II}; Cambridge University Press: Cambridge, UK, (1987).
\bibitem {stringbranerev} \Name{J.}{Polchinski} ~ TASI lecture on D-branes. \href{https://arxiv.org/abs/hep-th/9611050}{[arXiv:hep-th/9611050]}.
\bibitem {stringcond}  \Name {A.}{Adams} ; \Name {J.}{Polchinski} ; \Name {E.}{Silverstein} ~ Don't Panic! Closed String Tachyons in ALE Spacetimes. \Journal{\JHE}{0110}{2001}{029}, \href{https://arxiv.org/abs/hep-th/0108075}{[arXiv:hep-th/0108075]}.  
\bibitem {stringcond0} \Name{J.L.}{Karczmarek} ; \Name{A.}{Strominger} ~ Closed String Tachyon Condensation at c=1. \Journal{\JHE}{0405}{2004}{062}, \href{https://arxiv.org/abs/hep-th/0403169}{[arXiv:hep-th/0403169]}.
\bibitem {stringcond1} \Name{D.}{Green} ~ Nothing for Branes. \Journal{\JHE}{0704}{2007}{025}, \href{https://arxiv.org/abs/hep-th/0611003}{[arXiv:hep-th/0611003]}.
\bibitem {gaugestringcorr} \Name {S.S.}{Gubser} ; \Name {I.R.}{Klebanov} ; \Name {A.M.}{Polyakov} ~ Gauge Theory Correlators from Non-Critical String Theory. \Journal{\PLB}{428}{1998}{105}, \href{https://arxiv.org/abs/hep-th/9802109}{[arXiv:hep-th/9802109]}.
\bibitem {stringrev0} \Name {J.}{Polchinski} ~ \emph{String Theory I \& II}; Cambridge University Press: Cambridge, UK, 2005.
\bibitem {adscft5d} \Name{J.M.}{Maldacena} ; \Name{A.}{Strominger} ~ Semiclassical decay of near extremal fivebranes. \Journal{\JHE}{9712}{1997}{008}, \href{https://arxiv.org/abs/hep-th/9710014}{[arXiv:hep-th/9710014]}.
\bibitem {qgrmatrixu1} \Name{R.}{Bousso} ; \Name{A.L.}{Mints} ~ Holography and entropy bounds in the plane wave matrix model. \Journal {\PRD}{73}{2006}{126005}, \href{http://arxiv.org/abs/hep-th/0512201}{[arXiv:hep-th/0512201]}.
\bibitem {virasorosuinf} \Name{E.F.}{Floratos} ; \Name{J.}{Iliopoulos} ~ A Note on the Classical Symmetries of the Closed Bosonic Membranes. \Journal {\PLB}{201}{1988}{237}.
\bibitem {virasorosuinf0} \Name{I.}{Antoniadis} ; \Name{P.}{Ditsas} ; \Name{E.F.}{Floratos} ; \Name{J.}{Iliopoulos} ~ New Realizations of the Virasoro Algebra as Membrane Symmetries. \Journal {\NPB}{300}{1988}{549}.
\bibitem {stringgas} \Name{A.}{Nayeri} ; \Name{R.H.}{Brandenberger} ; \Name{C.}{Vafa} ~ Producing a Scale-Invariant Spectrum of Perturbations in a Hagedorn Phase of String Cosmology. \Journal {\PRL}{97}{2006}{021302}, \href{https://arxiv.org/abs/hep-th/0511140}{[arXiv:hep-th/0511140]}.
\bibitem {qgrmatrixrev} \Name{A.}{Konechny}, \Name{A.}{Schwarz} ~ Introduction to M(atrix) theory and noncommutative geometry. \Journal{\PRE}{360}{2002}{353}, \href{http://arxiv.org/abs/hep-th/0012145}{[arXiv:hep-th/0012145]}.
\bibitem {qgrmatrixrev0} \Name{H.}{Steinacker} ~ Emergent Geometry and Gravity from Matrix Models: an Introduction. \Journal {\CQG}{27}{2010}{133001}, \href{https://arxiv.org/abs/1003.4134}{[arXiv:1003.4134]}.
\bibitem {qftlargen} \Name{G.}{'t Hooft} ~ A Planar Diagram Theory for Strong Interactions. \Journal{\NPB}{72}{1974}{461}.
\bibitem {qgrmatrix0} \Name{R.}{Dijkgraaf} ; \Name{E.}{Verlinde} ; \Name{H.}{Verlinde} ~ Matrix String Theory. \Journal{\NPB}{500}{1997}{43}, \href{http://arxiv.org/abs/hep-th/9703030}{[arXiv:hep-th/9703030]}.
\bibitem {qgrmatrixthermal} \Name{N.}{Kawahara} ; \Name{J.}{Nishimura} ; \Name{S.}{Takeuchi} ~ High temperature expansion in supersymmetric matrix quantum mechanics. \Journal{\JHE}{0712}{2007}{103}, \href{https://arxiv.org/abs/0710.2188}{[arXiv:0710.2188]}.
\bibitem {schilgauge} \Name{A.}{Schild} ~ Classical null strings. \Journal{\PRD}{16}{1977}{1722}.
\bibitem {qgrmatrixnoncommut} \Name{A.}{Connes} ; \Name{M.R.}{Douglas} ; \Name{A.}{Schwarz} ~ Noncommutative Geometry and Matrix Theory: Compactification on Tori. \Journal{\JHE}{02}{1998}{003}, \href{http://arxiv.org/abs/hep-th/9711162}{[arXiv:hep-th/9711162]}.
\bibitem {qgrmatrixym} \Name{H.}{Steinacker} ~ Covariant Field Equations, Gauge Fields and Conservation Laws from Yang-Mills Matrix Models. \Journal {\JHE}{02}{2009}{044}, \href{https://arxiv.org/abs/0812.3761}{[arXiv:0812.3761]}.
\bibitem {landscapecount} \Name{J.}{Kumar} ~ A Review of Distributions on the String Landscape. \Journal {\IMA}{21}{2006}{3441}, \href{http://arxiv.org/abs/hep-th/0601053}{[arXiv:hep-th/0601053]}.
\bibitem {qgrmatrixinf} \Name{S.}{Brahma} ; \Name{R.}{Brandenberger} ; \Name{S.}{Laliberte} ~ Emergent Cosmology from Matrix Theory. \href{https://arxiv.org/abs/2107.11512}{[arXiv:2107.11512]}.
\bibitem {qgrmatrixinf0} \Name{H.}{Steinacker} ~ Gravity as a Quantum Effect on Quantum Space-Time. \href{https://arxiv.org/abs/2110.03936}{[arXiv:2110.03936]}.
\bibitem {qgrtest} The LIGO-Virgo Collaboration ~ Tests of General Relativity with Binary Black Holes from the second LIGO-Virgo Gravitational-Wave Transient Catalog. \Journal{\PRD}{103}{2021}{122002}, \href{https://arxiv.org/abs/2010.14529}{[arXiv:2010.14529]}.
\bibitem {manydodysymm} \Name{P.W.}{Anderson} ~ Absence of Diffusion in Certain Random Lattices. \Journal{\PRV}{109}{1958}{1492}.
\bibitem {manydodysymm0} \Name{T.}{Koma} ; \Name{H.}{Tasak} ~ Symmetry Breaking and Finite-Size Effects in Quantum Many-Body Systems. \Journal{\JSP}{76}{1994}{745}. 
\bibitem {houribranecr} \Name{H.}{Ziaeepour} ~ QCD Color Glass Condensate Model in Warped Brane Models. \Journal {\GCS}{11}{2005}{189}, \href{http://arxiv.org/abs/hep-ph/0412314}{[hep-ph/0412314]}.
\bibitem {seccurvproof} \Name{W.}{K\"uhnel} ~ Differential Geometry. Third Edition AMS, Rhode Island, (2010). 
\bibitem {seccurvproof0} \Name{J}{Gallier} ~ Differential Geometry and Lie Groups, Vol, I. Springer, (2020)
\end{thebibliography}
\end{document}